\documentclass[aps,prx,twocolumn,superscriptaddress,english,floatfix, longbibliography]{revtex4-2}

\usepackage{graphicx}
\usepackage{float}
\usepackage{physics}
\usepackage{cancel}
\usepackage{overpic}
\usepackage{enumerate}
\usepackage{textcomp}
\usepackage{amsmath}
\usepackage{amssymb}
\usepackage{pgf}
\usepackage{soul}
\usepackage[normalem]{ulem}
\usepackage{lipsum}
\usepackage{comment}
\usepackage{mathrsfs}
\usepackage{standalone}
\usepackage[colorlinks,citecolor=blue,linkcolor=blue,urlcolor=blue]{hyperref}
\usepackage{bm}
\usepackage{tikz}
\usepackage{orcidlink}
\usepackage{braket}
\usepackage{enumitem}
\usepackage{booktabs}
\usepackage[english,greek]{babel}
\usepackage{moreenum}
\usepackage{empheq}

\definecolor{darkgreen}{rgb}{0,0.5,0}

\graphicspath{{./}{./s/}{FigsZejian/}{FigsAnna/}{FigsGianluca/}}

\renewcommand{\selectlanguage}[1]{}

\begin{document}

\newcommand{\vpd}[0]{\vphantom{\dagger}}
\newcommand{\vps}[0]{\vphantom{*}}
\newcommand{\vpp}[0]{\vphantom{\prime}}

\newcommand{\hrho}[0]{\hat{\rho}}
\newcommand{\hrhoT}[0]{\hat{\rho}_{\text{T}}}
\newcommand{\Aop}[0]{\hat{A}^{\vpd}}
\newcommand{\Adag}[0]{\hat{A}^{\dagger}}
\newcommand{\ephi}[0]{e^{i\hat{\phi}}}
\newcommand{\enphi}[0]{e^{-i\hat{\phi}}}
\newcommand{\dJ}[0]{\delta\hat{J}}

\title{Generalized Holstein-Primakoff mapping and $1/N$ expansion of collective spin systems undergoing single particle dissipation}

\author{Diego Barberena~\orcidlink{0000-0002-6845-1807}}
%\author{Diego Barberena}
\affiliation{T.C.M. Group, Cavendish Laboratory, University of Cambridge, J.J. Thomson Avenue, Cambridge CB3 0US, UK}
% \author{Nigel Cooper~\orcidlink{0000-0002-4662-1254}}
% \affiliation{T.C.M. Group, Cavendish Laboratory, University of Cambridge, J.J. Thomson Avenue, Cambridge CB3 0HE, UK}
\begin{abstract}
    % We develop a generalization of the Schwinger boson and Holstein-Primakoff mappings that is applicable to spin $1/2$ systems undergoing single particle dissipation. We first write down an exact equivalence between generic terms of a spin master equation and an associated boson master equation written down in terms of Schwinger bosons. We then make contact with the original Holstein-Primakoff mapping by changing variables into a Holstein-Primakoff boson, which describes fluctuations transverse to the collective Bloch vector built out of the original spin $1/2$'s, and a longitudinal boson that describes fluctuations parallel to the Bloch vector. Using this representation, we develop a systematic $1/N$ expansion, for which we write down explicitly the leading and next-to-leading order terms. We then illustrate how to apply these results using three example systems: an ensemble of atoms undergoing spontaneous emission, incoherent pumping and single particle dephasing; a superradiant laser above and in the vicinity of the lasing transition; and the all-to-all transverse field Ising model subject to incoherent pumping in the vicinity of its phase transition.

     We develop a generalization of the Schwinger boson and Holstein-Primakoff transformations that is applicable to ensembles of $N$ spin $1/2$'s with weak permutational symmetry.  These generalized mappings are constructed by introducing two independent bosonic variables that describe fluctuations parallel and transverse to the collective Bloch vector built out of the original spin $1/2$'s. Using this representation, we develop a systematic $1/N$ expansion and write down explicitly leading and next-to-leading order terms. We then illustrate how to apply these techniques using four example systems: (i) an ensemble of atoms undergoing spontaneous emission, incoherent pumping and single particle dephasing; (ii) a superradiant laser above and in the vicinity of the upper lasing transition; (iii) the all-to-all transverse field Ising model subject to incoherent pumping in the vicinity of its ordering phase transition; and (iv) the Dicke model at finite temperature both away and in the vicinity of its thermal phase transition. Thus, these mappings provide a common, Bloch-sphere based, geometrical description of all-to-all systems subject to single particle dissipation or at finite temperature, including their phase transitions.

\end{abstract}
\maketitle

\section{Introduction}

Collective spin systems arise very frequently in the field of quantum technologies, e.g. when atoms interact with light inside an optical cavity~\cite{Mivehvar02012021} or when ions communicate via a common motional mode~\cite{Britton2012,Zhang2017DPT}. They often provide minimal theoretical descriptions of non-equilibrium phenomena such as superradiance~\cite{Dicke1954,GROSS1982301}, driven-dissipative dynamics~\cite{HJCarmichael_1980,Walls1978,Morrison2008,Kessler2012,Lee2014,Iemini2018,Kirton_2018,Ferreira2019,Titum2020,Somech2024, Baumann2010,Klinder2015,Kroeze2018,Ferioli2023,Song2025} and novel kinds of lasing~\cite{Chen2009,Meiser2009,Bohnet2012,Kazakov2013,Norcia2016,Schaffer2020}, and are thus fundamental ingredients of many iconic models from quantum optics~\cite{HJCarmichael_1980,Hepp1973,Kitagawa1993,Morrison2008}. Moreover, collective spin systems are also well suited for the preparation of highly entangled spin squeezed states~\cite{Wineland1992,Kitagawa1993,Nori2011,Pezze2018,Schleier-Smith2010,Leroux2010,Hosten2016,Cox2016,gilmore2021,PedrozoPenafiel2020}, with current efforts now focusing on using them for the improvement of state-of-the-art sensors~\cite{Robinson2024,yang2025clockprecisionstandardquantum}.

When collective spin systems are built out of ensembles of atoms, physical processes cannot fundamentally distinguish between the atoms that partake in them. This indiscernibility is a crucial ingredient for the creation of the quantum-enhanced correlations that underpin metrological applications, and often takes the form of a mathematically exact permutational symmetry among the atoms. This is advantageous because theoretical analyses based on this symmetry are considerably simpler~\cite{GROSS1982301}, while often still capturing the qualitative properties of similar but less symmetric models~\cite{Norcia2018,Perlin2020,Bilitewski2021,Comparin2022,Franke2023,Young2023,Bornet2023,Eckner2023}. Even when typical sources of decoherence such as spontaneous emission into free space are included, a restricted amount of this permutational symmetry is retained~\cite{Chase2008,Xu2013,Hartmann2016}, although correlations are usually damaged as a result.

The main technical simplification brought about by permutational symmetry is a reduction of the space of quantum states that the system explores during its dynamics. The typical, exponentially large in $N$, Hilbert space of $N$ spins is brought down to a subspace whose size is polynomial in $N$. The exact degree of reduction will then depend on whether the symmetry acts in a strong or a weak sense~\cite{Buca_2012,Albert2014,Lieu2020}. If there are only coherent interactions, governed by a Hamiltonian, or the sources of dissipation are collective, e.g. by coupling the atoms to a lossy cavity mode, the symmetry will be realized strongly. In this case, the size of the relevant subspace will be $\sim N(N^2)$ for closed (open) system dynamics. In contrast, in the presence of single particle sources of dissipation such as spontaneous emission, the symmetry will be realized weakly. The evolution will then be inherently open and the subspace of explored density matrices will be of size $\sim N^3$~\cite{Chase2008,bargaiola201,Hartmann2016}. These reductions are routinely exploited in numerical simulations~\cite{Shammah2018}, although the $N^3$ scaling is still very limiting in practice. Further gains can be achieved using (stochastic) Monte Carlo wavefunction techniques~\cite{Zhang_2018}, at the expense of requiring averages (with low statistical uncertainty) over multiple repetitions of the stochastic evolution.

\begin{figure}
    \centering
    \includegraphics[width=0.98\linewidth]{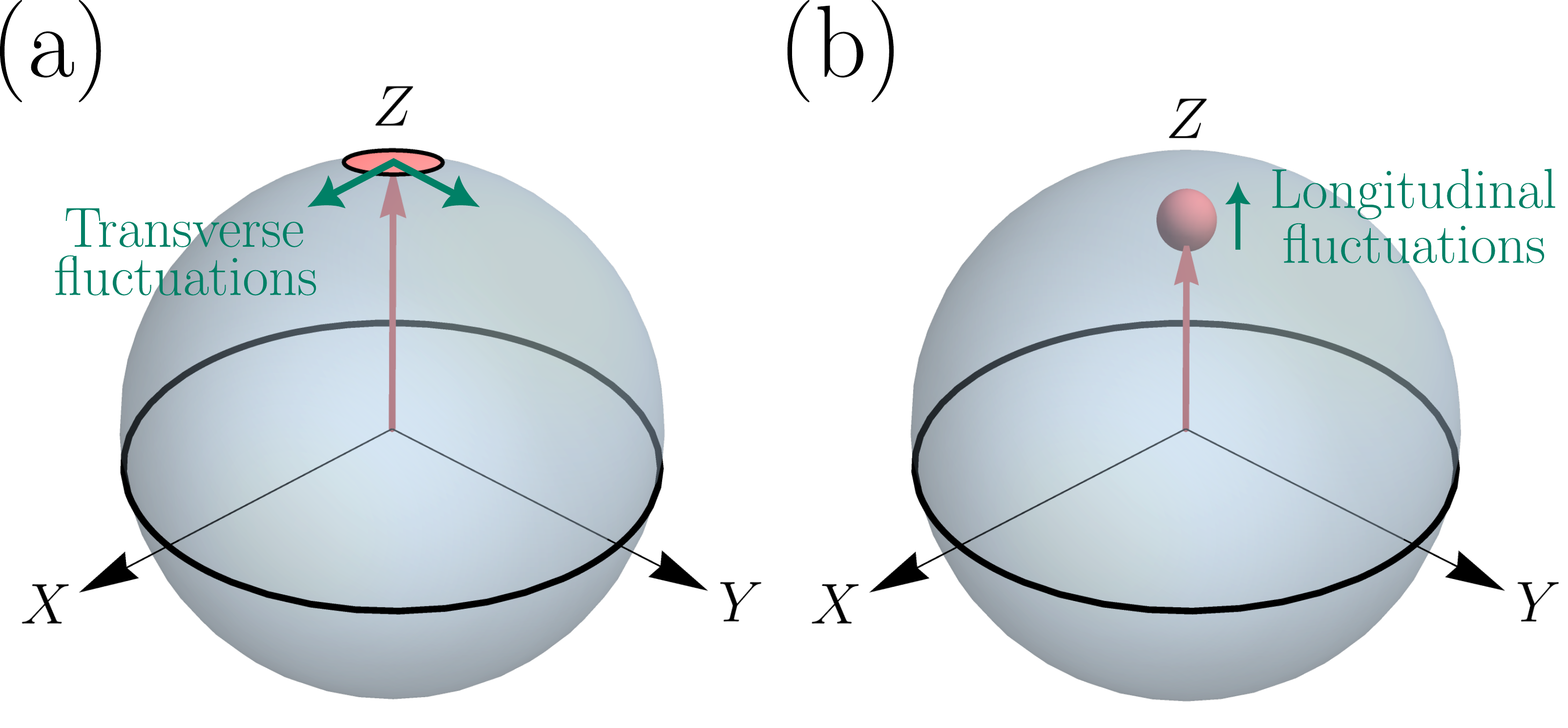}
    \caption{(a) For systems made up of $N$ spin $1/2$ and with strong permutational symmetry, both the collective Bloch vector (arrow in red) and fluctuations lie on the surface of the collective Bloch sphere, of radius $N/2$. Fluctuations are transverse to the collective Bloch vector and are described by the Holstein-Primakoff boson. (b) When the permutational symmetry is only weak, the Bloch vector can lie inside the sphere now, and there are also longitudinal fluctuations, which are described by a different boson. The noise distribution can now be three-dimensional.}
    \label{fig:Schematic}
\end{figure}

When the permutational symmetry is strong, analytical insight is often provided by using bosonic representations of the spin operators~\cite{Ressayre1975,hillery1985,Emary2003}. More concretely, collective spin operators can be expressed exactly in terms of two Schwinger bosons, in a way that makes manifest their nature as components of an $SO(3)$ vector. Moreover, the strongness of the symmetry provides a constraint that can be used to mathematically eliminate one of the bosons. The resulting expressions in terms of a single boson are known as the Holstein-Primakoff (HP) transformation~\cite{HolsteinPrimakoff1940}. Although HP hides manifest rotational covariance, it provides a way of performing a systematic expansion in powers of $1/N$~\cite{hillery1985,Emary2003}. The leading terms in this expansion give rise to the mean field approximation, and the leading corrections typically describe gaussian fluctuations about the mean field state. If the expansion is done carefully, it can also be used to analyze phase transitions, although in this case fluctuations acquire a non-gaussian character~\cite{Liberti2006,Liberti2010,Titum2020,Barberena2024}. Either way, the mean field quantum state can be visualized as an arrow (the Bloch vector) on the surface of a collective Bloch sphere, of radius $N/2$, while fluctuations can be represented as a small distribution about the tip of this arrow. Both the distribution and the arrow tip lie on the surface of the sphere [Fig.~\ref{fig:Schematic}(a)]. 

When the permutational symmetry is weak, i.e. in the presence of single particle decoherence or at finite temperature, the naive HP mapping breaks down and a Schwinger boson representation from which to obtain a modified HP approximation has not been derived in full generality. Applying second quantization in superoperator space leads to alternative bosonic representations~\cite{Sukharnikov2023} that recover the $N^3$ scaling, but the interpretation of the resulting bosons using Bloch spheres is not clear, and the nature of the large $N$ approximation in this representation has not been investigated. Following the more standard HP transformation, Ref.~\cite{KolmerForbes:24} used phase space methods and Fokker-Planck equations to analyse specific master equations. 

In this paper, we extend the results of Ref.~\cite{KolmerForbes:24} and provide a comprehensive description of generalized boson mappings for spin $1/2$ systems undergoing single particle dissipation. We identify a rotationally covariant structure that expresses local dissipation in terms of Schwinger bosons, and then use this representation to derive a modified HP transformation. We find that the original HP boson from strongly symmetric systems still appears and still describes fluctuations perpendicular to the mean field Bloch vector, which can now lie within the sphere and not only on its surface. In addition, there is now a second boson that accounts for longitudinal fluctuations parallel to the Bloch vector [Fig.~\ref{fig:Schematic}(b)]. 

This technique is of wide generality, as is the geometrical picture that accompanies it. To demonstrate this, this paper will fulfill three goals
\begin{enumerate}\label{intro:req}
    \item[($\alpha$)]Establish an exact operator mapping between generic terms in a spin Lindblad master equation and terms in an associated bosonic master equation.
    \item[($\beta$)]Devise a set of simple ``replacement rules" to analyze the large $N$ limit of such systems at their steady states, along with a clearly defined procedure on how to compute further corrections in $1/N$. When possible, these ``rules" should establish direct analogies to standard bosonic constructs (e.g. baths at some finite temperature).
    \item[($\Pi$)] Show that the exact mapping can also be used to get analytical control over the critical region of driven-dissipative and thermal phase transitions when the number of spins $N$ is sent to $\infty$.
\end{enumerate}
In carrying out items $(\alpha)$ and $(\beta)$, we will establish a simple ``recipe" to study generic all-to-all spin systems undergoing local dissipation or at finite temperature. Furthermore, item $(\Pi)$ is a generalization of techniques that have been used in the past to describe phase transitions of all-to-all systems~\cite{Liberti2006,Liberti2010,Titum2020} including only collective sources of dissipation~\cite{Barberena2024}. It also provides an alternative, operator based, analysis of phenomena that are more routinely studied using Keldysh path integral techniques~\cite{Sieberer_2016,DallaTorre2016,Paz2021}. To progressively fulfill the above three goals, we organize this paper as follows
\begin{description}
    \item[\underline{Section~\ref{sec:ModelsAndMotivation}}] \textit{Basic notation and review of $(\alpha),(\beta)$ and $(\Pi)$ for strong permutational symmetry.} We introduce the spin master equation that we will study and illustrate $(\alpha),(\beta)$ and $(\Pi)$ by focusing on the ground state properties of an example Hamiltonian.
    \item[\underline{Section~\ref{sec:DissipativeMapping}}] \textit{Items $(\alpha)$ and $(\beta)$ for weak permutational symmetry}. We write down the exact operator mappings and provide their large $N$ approximation. 
    \item[\underline{Section~\ref{sec:Examples}}] \textit{Examples}. We illustrate how to use the mappings away from phase transitions by means of two examples: a collection of spins undergoing single particle dephasing, spontaneous emission and incoherent pumping; and a superradiant laser above the upper lasing threshold.
    \item[\underline{Section~\ref{sec:PhaseTransitions}}] \textit{Item $(\Pi)$ for weak permutational symmetry}. We illustrate how to use the mappings in the vicinity of phase transition points by means of two examples: a superradiant laser near threshold and a driven-dissipative transverse field Ising model.
    \item[\underline{Section~\ref{sec:Thermal}}] \textit{Thermal states}. We show that the generalized mapping can also be used to analyze thermal properties of collective all-to-all models. We illustrate this using the Dicke model, and derive effective Hamiltonians in each of its two thermal phases and also in the vicinity of its thermal phase transition.
\end{description}
\section{Models and motivation}\label{sec:ModelsAndMotivation}
The type of models that we will study are defined in systems of $N$ qubits. The Hilbert space of each qubit is spanned by the states $\ket{\uparrow}_i$, $\ket{\downarrow}_i$, ($i=1,...,N$), in which local spin matrices 
\begin{align*}\begin{split}
    \hat{s}_x^i&=\frac{1}{2}(\ketbra{\uparrow}{\downarrow}_i+\ketbra{\downarrow}{\uparrow}_i)\\
    \hat{s}_y^i&=\frac{1}{2i}(\ketbra{\uparrow}{\downarrow}_i-\ketbra{\downarrow}{\uparrow}_i)\\
    \hat{s}_z^i&=\frac{1}{2}(\ketbra{\uparrow}{\uparrow}_i-\ketbra{\downarrow}{\downarrow}_i)
\end{split}\end{align*}
act. Importantly, the density matrix of the system $\hat{\rho}$ evolves under a Liouvillian with the following structure:
\begin{equation}\label{eqn:Models:MasterEquation}
    \partial_t\hat{\rho}=-i[\hat{H},\hat{\rho}]+\mathcal{L}\hat{\rho}+\sum_{\alpha,\beta,i}\gamma_{\alpha\beta}\left(\hat{s}_{\alpha}^i\hat{\rho}\hat{s}_{\beta}^i-\frac{\{\hat{s}_\beta^i\hat{s}_\alpha^i,\hat{\rho}\}}{2}\right),
\end{equation}
where $\hat{H}$ and $\mathcal{L}$ are a ``collective" Hamiltonian and Liouvillian, respectively, meaning that they are constructed entirely in terms of the collective spin operators $\hat{J}_{x,y,z}=\sum_i\hat{s}_{x,y,z}^i$ (and/or $\hat{J}^{\pm}=\hat{J}_x\pm i\hat{J}_y$). The third contribution, parameterized by the rates $\gamma_{\alpha\beta}$, describes single particle processes such as spontaneous emission, incoherent pumping, and dephasing. The collective parts, $\hat{H}$ and $\mathcal{L}$, describe instead processes such as superradiant emission of light (with jump operator $\propto\hat{J}^-$) or collective exchange interactions (with Hamiltonian $\propto \hat{J}^+\hat{J}^-$), which may arise via mediation of a cavity mode or a common motional mode. 

Collective spin operators are invariant under the action of permutation operators $\hat{U}_P$, i.e. $\hat{U}_P^\dagger\hat{J}_{x,y,z}\hat{U}_P=\hat{J}_{x,y,z}$. As a consequence, when $\gamma_{\alpha\beta}=0$ the evolution equation is independently invariant under $\hrho\to\hat{U}_P\hrho$ and $\hrho\to\hrho\hat{U}_P$. By definition, this means that permutations are a strong symmetry of the system~\cite{Buca_2012,Albert2014,Lieu2020}. It is then useful to construct the spin length operator $\hat{J}$, defined as the positive square root of
\begin{equation}\label{eqn:SpinLength}
    \hat{J}(\hat{J}+1)=\hat{J}_x^2+\hat{J}_y^2+\hat{J}_z^2.
\end{equation}
Using the eigenvalues of $\hat{J}$ as labels, we can then organize the $2^N$ possible quantum states of the system in terms of their behaviour under permutations. In particular, we will focus on the so-called Dicke manifold, which comprises all quantum states that are invariant under permutations. The dimension of the Dicke manifold for $N$ spins is $N+1$, and they are all eigenstates of $\hat{J}$ with eigenvalue $N/2$. A typical basis of this manifold is given by the Dicke states $\ket{J,M}$, which are also eigenstates of $\hat{J}_z$ with eigenvalue $M$ (we are keeping the label $J$ in the state to make connections with Sec.~\ref{sec:DissipativeMapping} more direct, although it has the value $N/2$ for the Dicke states).

\subsection{Motivation}\label{subsec:Motivation}
\begin{figure}
    \centering
    \includegraphics[width=0.98\linewidth]{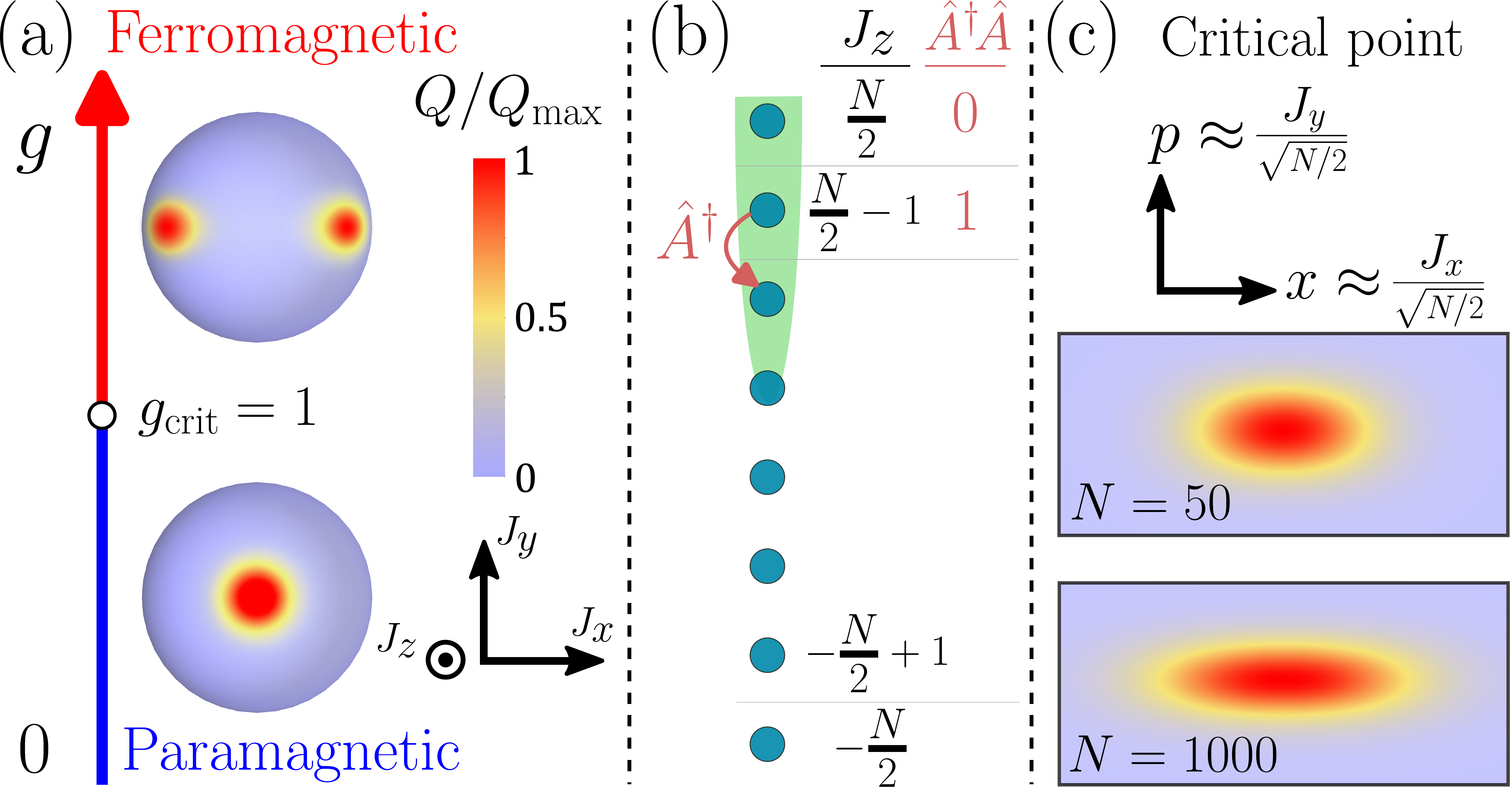}
    \caption{(a) Husimi distribution, $Q\propto |\braket{\theta,\phi|\text{gnd}}|^2$, of the ground state $\ket{\text{gnd}}$, where $\ket{\theta,\phi}$ is a spin-coherent state~\cite{JMRadcliffe_1971}. In the paramagnetic phase ($g<1$), the Bloch vector is polarized along $+z$. In the ferromagnetic ($g>1$) phase, the Bloch vector acquires a $\pm x$ component. (b) Dicke states, which are permutationally symmetric eigenstates of $\hat{J}_z$. The states can equally be labeled by the occupation number of the Holstein-Primakoff boson ($\Adag\protect\Aop$). The green shaded area is the region of Hilbert space where the large $N$ approximation is accurate. (c) Husimi distribution at the critical point $g=1$, plotted as a function of the quadratures $x\approx J_x/\sqrt{N/2}$ and $p\approx J_y/\sqrt{N/2}$ for $N=50,1000$. As $N$ increases the ground state gets squeezed along the $J_y$ direction.}
    \label{fig:Model:Motivaton:GroundState}
\end{figure}

To give a better characterization of what kind of description we are after, we will illustrate items $(\alpha),(\beta)$ and $(\Pi)$ of the introduction using the more familiar setting of Hamiltonian systems and ground states, so for now we set $\mathcal{L}=0$ and $\gamma_{\alpha\beta}=0$. We thus consider 
% a collection of $N$ spin $1/2$'s, described in terms of Pauli matrices $\hat{\sigma}_{x,y,z}^i$ ($i=1,...N$), which interact according to
an all-to-all version of the transverse field Ising model, also known as the Lipkin-Meshkov Glick model~\cite{LIPKIN1965188,Morrison2008,Morrison2008b}
\begin{equation}\label{eqn:Motivation:Hamiltonian}
    \hat{H}=-\hat{J}_z-\frac{g}{N}\hat{J}_x^2,
\end{equation}
which is expressed entirely in terms of collective spin operators. Because of the all-to-all connectivity of the model, mean field theory provides an accurate description of the ground state when $N \to\infty$. Specifically, there are two ground state phases [see Fig.~\ref{fig:Model:Motivaton:GroundState}(a)]:
\begin{itemize}
    \item Paramagnetic: When $g<1$, there is only one ground state, characterized by $\braket{\hat{J}_z}=N/2$ and $\braket{\hat{J}_{x,y}}=0$. Thus, the collective Bloch vector $\braket{\mathbf{\hat{J}}}=\braket{(\hat{J}_x,\hat{J}_y,\hat{J}_z)}$ points along the $+z$ direction.
    \item Ferromagnetic: When $g>1$, there are two degenerate ground states, with $\braket{\hat{J}_x}=\pm (N/2)\sqrt{1-1/g^2}$, $\braket{\hat{J}_y}=0$ and $\braket{\hat{J}_z}=N/(2g)$. The Bloch vector points now in the $xz$ plane.
\end{itemize}
Operationally, these are obtained by calculating the equations of motion for the expectations $\braket{\hat{J}_a}$, factorizing operator products $\braket{\hat{J}_a\hat{J}_b}\to\braket{\hat{J}_a}\braket{\hat{J}_b}$, setting the time derivatives to $0$ and solving the ensuing nonlinear equations. Alternatively, in the case of ground states the mean field solution can also be obtained by replacing operators by classical variables $(\hat{J}_x,\hat{J}_y,\hat{J}_z)\to N(\sin\theta\cos\phi,\sin\theta\sin\phi,\cos\theta)$ and minimizing the resulting classical Hamiltonian with respect to the parameters $\theta,\phi$.

Studying fluctuations requires going beyond the simple factorizaton scheme used to obtain the mean field results. This is achieved by representing the collective spin operators in terms of Schwinger bosons
\begin{equation}\label{eqn:Motivation:SchwingerBosons}
    \mathbf{\hat{J}}=\frac{1}{2}\begin{pmatrix}
        \hat{b}^\dagger&\hat{a}^\dagger
    \end{pmatrix}\,\bm{\sigma}\begin{pmatrix}
        \hat{b}\\ \hat{a}
    \end{pmatrix},
\end{equation}
where $\bm{\sigma}=(\sigma_x,\sigma_y,\sigma_z)^T$, and ($\hat{a},\hat{a}^\dagger$) and ($\hat{b},\hat{b}^\dagger$) are two pairs of bosonic variables satisfying standard commutation relations $[\hat{a},\hat{a}^\dagger]=[\hat{b},\hat{b}^\dagger]=1$. In the Schwinger boson representation, the spin length $\hat{J}$ also has a simple form
\begin{equation}
    \hat{J}=\frac{a^\dagger\hat{a}+\hat{b}^\dagger\hat{b}}{2}.
\end{equation}
For Hamiltonian systems, this constitutes item $(\alpha)$ in the introduction.

Physically, the number operators $\hat{b}^\dagger\hat{b}$ and $\hat{a}^\dagger\hat{a}$ count the number of spins in $\ket{\uparrow}$ and $\ket{\downarrow}$ respectively. Because of permutational symmetry, basis states in the Dicke manifold are specified uniquely by the occupation numbers of $\ket{\uparrow}$ and $\ket{\downarrow}$. Since there are $N$ spins in total, quantum states $\ket{\psi}$ in the Dicke manifold are subject to the constraint
\begin{equation}\label{eqn:Motivation:Constraint}
    (\hat{a}^\dagger\hat{a}+\hat{b}^\dagger\hat{b})\ket{\psi}=N\ket{\psi},
\end{equation}
or equivalently $\hat{J}\ket{\psi}=(N/2)\ket{\psi}$. To study fluctuations in the paramagnetic ground state, we recognize that the collective spin points along $+z$, so that $\hat{b}^\dagger\hat{b}\sim N$ and $\hat{a}^\dagger\hat{a}\sim 1$. It is thus convenient to use the number-phase representation for $\hat{b}=e^{i\hat{\phi}/2}(\hat{b}^\dagger\hat{b})^{1/2}$~\cite{Susskind1964}, where $e^{i\hat{\phi}/2}$ reduces the occupation of the $\hat{b}$ boson by $1$ with unit amplitude. Using the constraint Eq.~(\ref{eqn:Motivation:Constraint}) and defining $\hat{A}=\hat{a} e^{-i\hat{\phi}/2}$ enables us to represent the collective spin operators in terms of a single Holstein-Primakoff boson~\cite{HolsteinPrimakoff1940}
\begin{align}
    \begin{split}
        \hat{J}_z&=\frac{N}{2}-\hat{A}^\dagger\hat{A}\\
        \hat{J}^+&=(N-\hat{A}^\dagger\hat{A})^{1/2}\times \hat{A}\\
        \hat{J}^-&=\hat{A}^\dagger\left(N-\hat{A}^\dagger\hat{A}\right)^{1/2}
    \end{split}
\end{align}
This is an alternative version of item $(\alpha)$ in the introduction. The Holstein-Primakoff mapping is exact, but it is more convenient when the Bloch vector is aligned along $+z$, because then $\braket{\hat{A}^\dagger\hat{A}}\ll N$, $\text{Var}(\hat{A}^\dagger\hat{A})\ll N^2$, and the square roots can be expanded in a Taylor series. This establishes a systematic way of studying fluctuations with a controlled small parameter $N^{-1/2}$. For later convenience, we also define here the quadrature operators $\hat{x}=(\hat{A}+\hat{A}^\dagger)/\sqrt{2}$ and $\hat{p}=-i(\hat{A}-\hat{A}^\dagger)/\sqrt{2}$.

To leading order in $1/N$, we can approximate 
\begin{align}
    \begin{split}
        \hat{J}^+&\approx \sqrt{N}\hat{A}\\
        \hat{J}^{-}&\approx \sqrt{N}\hat{A}^\dagger\\
        \hat{J}_z&=N/2-\hat{A}^\dagger\hat{A}
    \end{split}
\end{align}
This set of replacement rules constitutes item $(\beta)$ in the introduction, and provides a direct analogy to boson creation/destruction processes. Replacing these expressions in Eq.~(\ref{eqn:Motivation:Hamiltonian}) leads to the fluctuation Hamiltonian
\begin{equation}\label{eqn:Motivation:HamiltonianFluctuations}
    \hat{H}\approx -\frac{(N+1)}{2}+\frac{\hat{p}^2}{2}+\frac{(1-g)\hat{x}^2}{2}+O(N^{-1/2}).
\end{equation}
Since the Hamiltonian is quadratic in boson operators, expectation values can be calculated analytically. For instance,
\begin{align}\label{eqn:Motivation:Fluctuations}
    \begin{split}
        \braket{\hat{J}_x^2}&\approx \frac{N}{4}(1-g)^{-1/2}\\
        \braket{\hat{J}_y^2}&\approx  \frac{N}{4}(1-g)^{1/2}\\
    \end{split}
\end{align}

To analyze fluctuations about the ferromagnetic ground state, one first needs to rotate the mean field collective Bloch vector (which is now tilted in the $xz$ plane) to the $+z$ axis. All the other steps then follow through identically.

As we approach the critical point through the paramagnetic phase ($g\to 1^-$), fluctuations in $\hat{J}_x\sim\hat{x}$ diverge. This just means that the leading order approximation in $1/N$ fails, but the Holstein-Primakoff mapping remains exact. To get control over the phase transition region, we need to keep the relevant nonlinearity in the next order in $1/N$, which is given by
\begin{equation}\label{eqn:Models:Motivation:CorrectedJx}
    \hat{J_x}\approx \sqrt{\frac{N}{2}}\,\hat{x}-\frac{\hat{x}^3}{4\sqrt{2N}}.
\end{equation}
Note that, at the same order in $1/N$, there are also terms of the form $\hat{x}^2\hat{p}$, but these are smaller than $\hat{x}^3$ on account of $\hat{x}$ being the variable with diverging fluctuations. The corrected Hamiltonian near $g=1$ is thus
\begin{equation}
    \hat{H}\approx -\frac{(N+1)}{2}+\frac{\hat{p}^2}{2}+\frac{(1-g)\hat{x}^2}{2}+\frac{\hat{x}^4}{4N}.
\end{equation}
The dependence with $N$ can be made manifest by canonically rescaling $\hat{x}=N^{1/6}\hat{y}$ and $\hat{p}=N^{-1/6}\hat{q}$, and introducing a scaled coupling constant $\xi$ that measures deviations from the critical point according to $g=1-\xi/N^{2/3}$. In terms of these variables $\hat{H}$ becomes
\begin{equation}\label{eqn:Models:NonlinearHammiltonianPhaseTransiton}
    \hat{H}\approx -\frac{(N+1)}{2}+\frac{1}{N^{1/3}}\left(\frac{\hat{q}^2}{2}+\frac{\xi\hat{y}^2}{2}+\frac{\hat{y}^4}{4}\right),
\end{equation}
which is still a bosonic Hamiltonian, but now with a non-gaussianity that is especially relevant at $g=1$ and an energy gap that scales like $N^{-1/3}$~\cite{Dusuel2004}. Using this formulation, and since $\hat{y},\hat{q}\sim 1$, we immediately see that fluctuations in $\hat{J}_y$ at the critical point scale as $\hat{J}_y\sim\sqrt{N}\hat{p}\sim N^{1/3}$ and so the state is squeezed along the $y$ direction. Because of the same reason, the terms neglected in Eq.~(\ref{eqn:Models:Motivation:CorrectedJx}) are of size $\hat{x}^2\hat{p}\sim N^{1/6}$ and parametrically smaller than $ \hat{x}^3\sim N^{1/2}$, justifying their omission a posteriori, and indicating that the large $N$ expansion in the vicinity of the critical point is in fact an expansion in powers of $N^{1/3}$ rather than $N^{1/2}$. This type of analysis~\cite{Liberti2006,Liberti2010,Titum2020,Kirton_2018,DallaTorre2016} constitutes item ($\Pi$) in the introduction. \\

\section{Operator mapping}\label{sec:DissipativeMapping}
We now consider all of Eq.~(\ref{eqn:Models:MasterEquation}), following closely the logic of Sec.~\ref{subsec:Motivation}, and begin by discussing permutational symmetry. Unlike collective Hamiltonians and Liouvillians, local dissipation is only symmetric with respect to permutations in the weak sense~\cite{Buca_2012,Albert2014,Lieu2020}. In consequence, Eq.~(\ref{eqn:Models:MasterEquation}) does not preserve the Dicke manifold, and the associated steady state Bloch vector is no longer restricted to have the maximum length $N/2$. Nevertheless, it is still possible to define a ``generalized" Dicke manifold~\cite{Chase2008,Hartmann2016} of density matrices, which IS preserved by Eq.~(\ref{eqn:Models:MasterEquation}), and which can still be depicted using Bloch spheres [see Fig~\ref{fig:Schematic}(b)].
\begin{figure}
    \centering
    \includegraphics[width=0.65\linewidth]{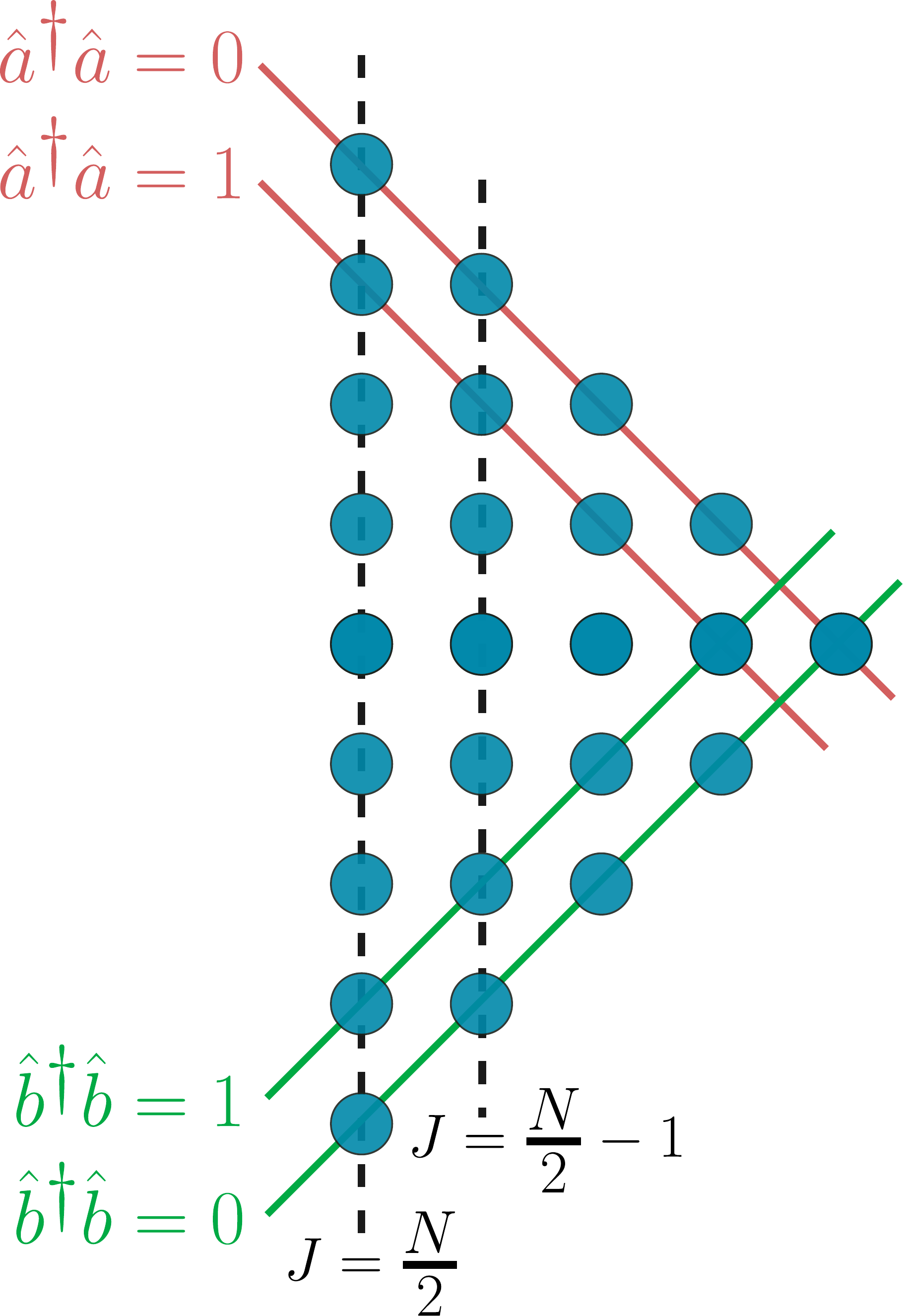}
    \caption{Dicke triangle, with ``states" $\ket{J,M}$, and their enumeration in terms of Schwinger bosons. The column with $J=N/2$ corresponds to the Dicke states from Fig.~\ref{fig:Model:Motivaton:GroundState}(b).}
    \label{fig:DissipativeHP:BlochSpheres}
\end{figure}

The ``generalized" Dicke manifold for $N$ spins is spanned by the matrices $\overline{\ketbra{J,M}{J,M'}}$, which are the unique (up to normalization) permutationally symmetric density matrices that are also
% \begin{align}\begin{split}\label{eqn:DHP:GeneralizedDicke}
%     \hat{J}\hspace{0.2cm}\overline{\ketbra{J,M}{J,M'}}&=\overline{\ketbra{J,M}{J,M'}}\hspace{0.2cm}\hat{J}=J \hspace{0.2cm}\overline{\ketbra{J,M}{J,M'}}\\
%     \hat{J}_z\hspace{0.2cm}\overline{\ketbra{J,M}{J,M'}}&=M\hspace{0.2cm}\overline{\ketbra{J,M}{J,M'}}\\
%     \overline{\ketbra{J,M}{J,M'}}\hspace{0.2cm}\hat{J}_z&=M'\hspace{0.2cm}\overline{\ketbra{J,M}{J,M'}}.
% \end{split}\end{align}
right/left eigenstates of $\hat{J}$ with equal eigenvalue $0<J\leq N/2$, and right/left eigenstates of $\hat{J}_z$ with eigenvalues $M,M'$, respectively. Importantly, the $\overline{\ketbra{J,M}{J,M'}}$ is not an outer product of Dicke states with different $J,M$, but in many respects it behaves like one, so it is useful to picture the dynamics of the system as if it were happening in a Hilbert space spanned by the states $\ket{J,M}$. This defines the Dicke triangle~\cite{Shammah2017,Zhang_2018,Debnath2018,Wu2024}, depicted in Fig.~\ref{fig:DissipativeHP:BlochSpheres}(b). The upshot of all of this is that the $4^N$ dimensional space of density matrices is reduced to a subspace of dimension $\sim N^3$. 

Given Eq.~(\ref{eqn:Models:MasterEquation}), the first line of attack is a mean field analysis. We assume that this has been done, resulting in a mean field Bloch vector $\mathbf{J}_{\text{mf}}=(J_x^{\text{mf}},J_y^{\text{mf}},J_z^{\text{mf}})$, and that the axes have been rotated so that $\mathbf{J}_{\text{mf}}$ is aligned with the positive $z$ axis. We thus have that $J_x^{\text{mf}}=J_y^{\text{mf}}=0$ and $J_z^{\text{mf}}>0$. Moreover, the length of the Bloch vector $J_{\text{mf}}$ coincides with $J_{z}^{\text{mf}}$.

To study fluctuations, we will make use of the Schwinger boson representation in Eq.~(\ref{eqn:Motivation:SchwingerBosons}), which we reproduce here for reference purposes
\begin{align}\begin{split}\label{eqn:DHP:SchwingerBosons}
    \mathbf{\hat{J}}&=\frac{1}{2}\begin{pmatrix}
        \hat{b}^\dagger&\hat{a}^\dagger
    \end{pmatrix}\,\bm{\sigma}\begin{pmatrix}
        \hat{b}\\ \hat{a}
    \end{pmatrix}\\[5pt]
    \hat{J}&=\frac{\hat{a}^\dagger\hat{a}+\hat{b}^\dagger\hat{b}}{2}.
\end{split}\end{align}
The main difference with respect to Sec.~\ref{subsec:Motivation} is that $2\hat{J}=\hat{a}^\dagger\hat{a}+\hat{b}^\dagger\hat{b}$ is now allowed to fluctuate. Since basis elements of density matrices in Schwinger boson space are specified by four numbers (left and right eigenvalues of $\hat{a}^\dagger\hat{a}$ and $\hat{b}^\dagger\hat{b}$), they can accommodate the three index object $\overline{\ketbra{J,M}{J,M'}}$ [see Fig.~\ref{fig:DissipativeHP:BlochSpheres}].

The Hamiltonian, the collective Liouvillian, and the anticommutator terms of Eq.~(\ref{eqn:Models:MasterEquation}) are constructed in terms of collective operators, so for them Eq.~(\ref{eqn:DHP:SchwingerBosons}) suffices. However, terms such as $\hat{s}_\alpha\hat{\rho}\,\hat{s}_\beta$ require a distinct bosonic description. We construct it by combining rotational properties of the spin operators with the results from Ref.~\cite{Chase2008}, which provides the (superoperator) matrix elements of $\hat{s}_\alpha\hat{\rho}\,\hat{s}_\beta$ between generalized Dicke states. To do this, we recall that spin operators transform as $SO(3)$ vectors under rotations, a feature that is made manifest in Eq.~(\ref{eqn:DHP:SchwingerBosons}), given that $(\hat{b}\,\hat{a})^T$ transforms as a $SU(2)$ doublet. There are, however, more ways of constructing $SO(3)$ vectors out of $(\hat{b}\,\hat{a})^T$. For local dissipation, we will need
\begin{align}
    \begin{split}
        \mathbf{\hat{K}}&=\frac{1}{2}\begin{pmatrix}
        \hat{b}&\hat{a}
    \end{pmatrix}\,i\sigma_y\bm{\sigma}\begin{pmatrix}
        \hat{b}\\ \hat{a}
    \end{pmatrix}\\[5pt]
    \mathbf{\hat{L}}&=-\frac{1}{2}\begin{pmatrix}
        \hat{b}^\dagger&\hat{a}^\dagger
    \end{pmatrix}\bm{\sigma}\,i\sigma_y\begin{pmatrix}
        \hat{b}^\dagger\\ \hat{a}^\dagger
    \end{pmatrix}=\mathbf{\hat{K}}^\dagger.
    \end{split}
\end{align}
The vector $\mathbf{\hat{K}}$ ($\mathbf{\hat{L}}$) is constructed out of two destruction (creation) operators, so it changes the value of $\hat{J}$ by $-1$ ($+1$). Using the three vectors $\mathbf{J},\mathbf{K},\mathbf{L}$, we can express local dissipator terms as (see Appendix~\ref{app:SchwingerDissipators})
\begin{equation}\label{eqn:DH:GeneralizedSchwinger}
    \sum_{i=1}^N\hat{s}_{\alpha}^i\hat{\rho}\,\hat{s}^i_{\beta}=\hat{E}\,\hat{J}_\alpha\hat{\rho}\hat{J}_\beta+\hat{F}\,\hat{K}_\alpha\hat{\rho}\hat{L}_\beta+\hat{G}\,\hat{L}_\alpha\hat{\rho}\hat{K}_\beta,
\end{equation}
where
\begin{align}\begin{split}
    \hat{E}&=\frac{1+N/2}{2\hat{J}(\hat{J}+1)}\\[5pt]
    \hat{F}&=\frac{N/2+\hat{J}+2}{2(\hat{J}+1)(2\hat{J}+3)}\\[5pt]
    \hat{G}&=\frac{N/2-\hat{J}+1}{2\hat{J}(2\hat{J}-1)}
\end{split}\end{align}
are functions only of the spin length $\hat{J}$. At this level of generality, this corresponds to item $(\alpha)$ in the introduction. 

To build up a Holstein-Primakoff mapping, we need to get rid of the $\hat{b}$ boson. As in Sec.~\ref{subsec:Motivation}, we do this by using a number-phase decomposition for $\hat{b}=e^{i\hat{\phi}/2}(\hat{b}^\dagger\hat{b})^{1/2}$, introducing $\hat{A}=\hat{a} e^{-i\hat{\phi}/2}$, and replacing $\hat{b}^\dagger\hat{b}$, wherever it appears, using the relation $\hat{b}^\dagger\hat{b}=2\hat{J}-\hat{A}^\dagger\hat{A}$. We also introduce $\delta\hat{J}=\hat{J}-J_{\text{mf}}$, which measures fluctuations of $\hat{J}$ with respect to its mean field value. As a consequence of these choices, both $\delta\hat{J}$ and $\hat{A}^\dagger\hat{A}$ are $\ll N$. This furnishes two independent, physically transparent, sets of variables:
\begin{itemize}
    \item The pair $\dJ, \ephi$ describes fluctuations parallel to the mean field spin direction, and satisfies a standard number/phase relation $\dJ \ephi=\ephi(\dJ-1)$. Note that $e^{i\hat{\phi}}$ reduces $\dJ$ by 1 while keeping $\Adag\Aop$ fixed, so it reduces $\hat{b}^\dagger\hat{b}$ by $2$.
    \item The pair $\Aop,\Adag$ satisfy $[\Aop,\Adag]=1$ and describe fluctuations transverse to the mean field spin direction. This is most easily seen by considering $\hat{J}^+=\hat{a}\hat{b}^\dagger\approx \Aop\sqrt{2J_{\text{mf}}}$. The prefactor in front of $\Aop$ is no longer exactly $\sqrt{N}$ but is still of the same order since $J_{\text{mf}}\sim N$.
\end{itemize}

\begin{figure}
    \centering
    \includegraphics[width=0.98\linewidth]{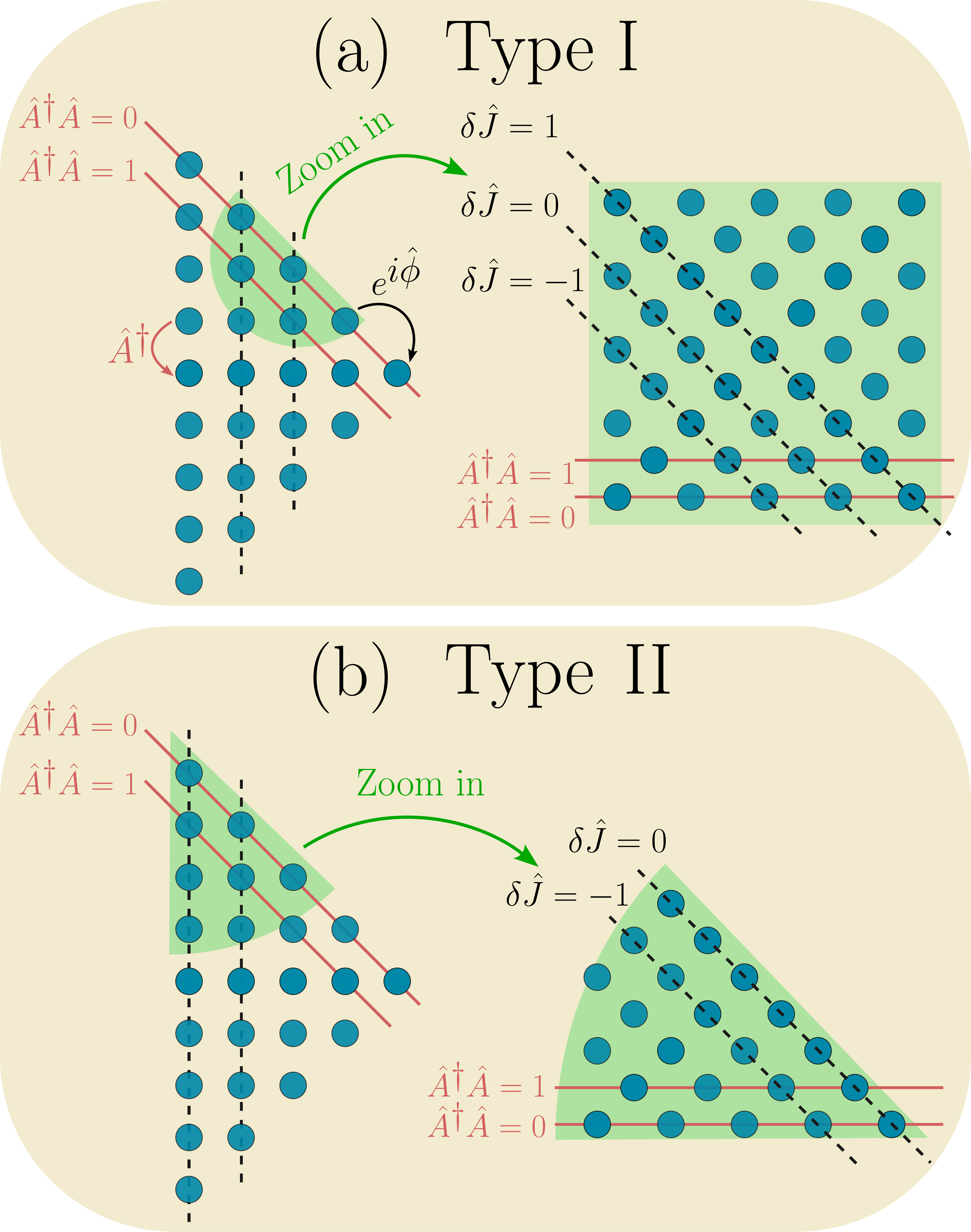}
    \caption{(a) Dicke triangle, now described in terms of the Holstein-Primakoff boson and $\hat{J}$. The operator $\protect\Aop$ moves vertically, while $\ephi$ moves parallel to upper boundary. Shaded green region represents states around a type I steady state, i.e., polarized along $+z$ with mean field length $J_{\text{mf}}<N/2$. The large $N$ expansion focuses on states within the shaded region, which upon zooming in becomes a half-plane ($\Adag\protect\Aop>0$, and no restriction on $\dJ$). (b) Dicke triangle again, but shaded region now represents states around a type II steady state, i.e., polarized along $+z$ with mean field length $J_{\text{mf}}= N/2$. The shaded region is now a squashed quarter plane, with $\Adag\protect\Aop>0$ and $\dJ<0$.}
    \label{fig:DissipativeHP:HP}
\end{figure}
From this representation we can now perform a large $N$ approximation systematically. The qualitative nature of the expansion will depend on whether the (mean field) normalized spin length $j=J_{\text{mf}}/(N/2)$ is less than or equal to $1$. We treat these cases independently and call them type I and type II, respectively, for ease of reference.
\subsection{Type I: Replacement rules when $j<1$}
When $j<1$, the mean field steady state is localized along the upper boundary of the Dicke triangle, but away from the corners, as depicted by the shaded region in Fig.~\ref{fig:DissipativeHP:HP}(a). The fluctuations in $\delta\hat{J}$ will generically be of size $\sqrt{N}$, while $\hat{\phi}$ can be taken to be sharply defined, with fluctuations $\delta\hat{\phi}\sim 1/\sqrt{N}$. This means that $e^{i\hat{\phi}}$ can be expanded in a Taylor series. To take this into account, and to make manifest the various scalings with $N$, we introduce normalized longitudinal bosons $\hat{l}=N^{-1/2}\delta\hat{J}$, $\hat{q}=\sqrt{N}\hat{\phi}$, with commutator $[\hat{l},\hat{q}]=i$. After some algebra (see Appendix~\ref{app:ReplacementRuleDerivation}), we obtain the associated bosonic approximations of the spin dissipators, as shown in Table~\ref{tab:typLind}.
\begin{table*}[h!]
\centering
\setlength{\tabcolsep}{12pt}
\begin{tabular}{c||c}\toprule
 Spin term &Boson term\\[5pt] \hline\hline
$\sum_{i=1}^N\left(\hat{s}_z^i\hat{\rho}\hat{s}_z^i-\frac{1}{2}\{\hat{s}_z^i\hat{s}_z^i,\hat{\rho}\}\right)$
&\parbox{4cm}{\begin{align*} \begin{split}
     &\left(\frac{j+1}{2j}\right) \left(\hat{A} \hat{\rho}\hat{A}^{\dagger} -\frac{1}{2}\{\hat{A}^{\dagger}\hat{A},\hat{\rho}\}\right)+\left(\frac{1-j}{2j}\right)\left(\hat{A}^{\dagger}\hat{\rho}\hat{A} -\frac{1}{2}\{\hat{A}\hat{A}^\dagger,\hat{\rho}\}\right)
 \end{split}\end{align*}}
\\[5pt] \hline

$\sum_{i=1}^N\left(\hat{\sigma}_+^i\hat{\rho}\hat{\sigma}_-^i-\frac{1}{2}\{\hat{\sigma}_-^i\hat{\sigma}_+^i,\hat{\rho}\}\right)$
&\parbox{4cm}{\begin{align*} \begin{split}
    &\frac{1}{j}\left(\hat{A}\hat{\rho}\hat{A}^{\dagger}-\frac{1}{2}\{\hat{A}^{\dagger}\hat{A},\hat{\rho}\}\right)+\frac{(1-j)}{4}\big[\hat{q},\big[\hat{\rho},\hat{q}\big]\big]+\frac{i}{2}\left[\hat{q},\{\hat{l},\hat{\rho}\}\right]-\frac{i\sqrt{N}(1-j)}{2}\left[\hat{q},\hat{\rho}\right]
 \end{split}\end{align*}}\\[5pt]\hline
$\sum_{i=1}^N\left(\hat{\sigma}_-^i\hat{\rho}\hat{\sigma}_+^i-\frac{1}{2}\{\hat{\sigma}_+^i\hat{\sigma}_-^i,\hat{\rho}\}\right)$
&\parbox{4cm}{\begin{align*} \begin{split}
    &\frac{1}{j}\left(\hat{A}^\dagger\hat{\rho}\hat{A}-\frac{1}{2}\{\hat{A}\hat{A}^\dagger,\hat{\rho}\}\right)+\frac{(1+j)}{4}\big[\hat{q},\big[\hat{\rho},\hat{q}\big]\big]+\frac{i}{2}\left[\hat{q},\{\hat{l},\hat{\rho}\}\right]+\frac{i\sqrt{N}(1+j)}{2}\left[\hat{q},\hat{\rho}\right]
 \end{split}\end{align*}}\\[5pt]
\hline\hline
$\sum_{i=1}^N\left(\hat{\sigma}_+^i\hat{\rho}\hat{\sigma}_+^i-\frac{1}{2}\{\hat{\sigma}_+^i\hat{\sigma}_+^i,\hat{\rho}\}\right)$
&\parbox{4cm}{\begin{align*} \begin{split}
   \frac{1}{j} \left(\hat{A}\hat{\rho}\hat{A}-\frac{1}{2}\{\hat{A}^2,\hat{\rho}\}\right)-\frac{1}{2}\left[\hat{A}^2,\hat{\rho}\right]
 \end{split}\end{align*}}\\[5pt]
\hline
 $\sum_{i=1}^N\left(\hat{\sigma}_+^i\hat{\rho}\hat{s}_z^i-\frac{1}{2}\{\hat{s}_z^i\hat{\sigma}_+^i,\hat{\rho}\}\right)$
&\parbox{4cm}{\begin{align*} \begin{split}
  \sqrt{N}\left(\frac{2-j}{4\sqrt{j}}\right)\left[\hat{A},\hat{\rho}\right]-\left(\frac{2+j}{4j\sqrt{j}}\right)\left[\Aop,\hat{\rho}\right]\hat{l}+i\left(\frac{1-j}{2\sqrt{j}}\right)\left[\hat{q},\hat{\rho}\right]\hat{A}
 \end{split}\end{align*}}\\[5pt]\hline
 $\sum_{i=1}^N\left(\hat{\sigma}_-^i\hat{\rho}\hat{s}_z^i-\frac{1}{2}\{\hat{s}_z^i\hat{\sigma}_-^i,\hat{\rho}\}\right)$
&\parbox{4cm}{\begin{align*} \begin{split}
  \sqrt{N}\left(\frac{j+2}{4\sqrt{j}}\right)\left[\Adag,\hat{\rho}\right]-\left(\frac{2-j}{4j\sqrt{j}}\right)\left[\Adag,\hat{\rho}\right]\hat{l}-i\left(\frac{1+j}{2\sqrt{j}}\right)\left[\hat{q},\hat{\rho}\right]\hat{A}^\dagger
 \end{split}\end{align*}}\\[5pt]\hline
\bottomrule
\end{tabular}
\caption{Replacement rules, to order $N^0$, when $j<1$. The first three lines correspond to typical Lindbladian terms that describe white-noise dephasing (first line), incoherent pumping  (second line) and incoherent decay (third line). The remaining dissipators can be obtained by conjugation of the last three lines.}\label{tab:typLind}
\end{table*}

These ``replacement rules" constitute item $(\beta)$ in the introduction. At the same time, they provide intuitive bosonic pictures. For example, in white-noise dephasing ($\hat{s}_z\hat{\rho}\,\hat{s}_z$) the transverse boson behaves as if it were connected to a finite temperature bath with absorption/emission rates that depend on the normalized mean field spin length $j$. In incoherent pumping ($\hat{\sigma}_+\hat{\rho}\,\hat{\sigma}_-$), which drives the system towards $+z$, the transverse boson is connected to a $0$ temperature bath, while the longitudinal boson ($\hat{l}$) is subject to diffusion (term $[\hat{q},[\hat{\rho},\hat{q}]]$) and relaxation (term $[\hat{q},\{\hat{l},\hat{\rho}\}]$). In incoherent decay ($\hat{\sigma}_-\hat{\rho}\,\hat{\sigma}_+$), which drives the Bloch vector towards $-z$ and away from $+z$, the transverse boson is instead connected to a $\infty$ temperature bath. Note also that some dissipators (including incoherent pumping/decay) have terms that are proportional to $\sqrt{N}$. These terms should cancel in the full Liouvillian (only) if the expansion is done about the correct mean field steady state.

\subsection{Type II: Replacement rules when $j=1$}
When $j=1$, the mean field steady state is localized near the upper corner of the Dicke triangle, as depicted by the shaded region in Fig.~\ref{fig:DissipativeHP:HP}(b). Now $\delta\hat{J}<0$ (the spin length can only be smaller than $N/2$) and fluctuations will also be $\delta\hat{J}\sim 1$. Consequently, the phase variable will have fluctuations $\delta\hat{\phi}\sim 1$ and we can no longer Taylor expand the exponential. Instead, we have to keep the expressions for the longitudinal boson intact. Physically, this means that the discreteness of $\delta\hat{J}$ is relevant. This leads to the replacement rules shown in Table~\ref{tab:typLindII} (see Appendix~\ref{app:ReplacementRuleDerivation}).

\begin{table*}[h!]
\centering
\setlength{\tabcolsep}{10pt}
\begin{tabular}{c||c}\toprule 
 Spin term &Boson term\\[5pt] \hline\hline
$
\sum_{i=1}^N\hat{s}_z^i\hrho\hat{s}_z^i
$
%%%%%%%%
&\parbox{8cm}{
\begin{equation*}
\frac{N\hrho}{4} +\Aop e^{i\hat{\phi}}\hrho \,e^{-i\hat{\phi}}\Adag-\frac{1}{2}\{\Adag\Aop,\hrho\}+O(N^{-1})
\end{equation*}
}\\ \hline
%%%%%%%%
$\sum_{i=1}^N\hat{\sigma}_+^i\hrho\hat{\sigma}_-^i$
&\parbox{8cm}{\begin{equation*}\Aop\hat{\rho}\Adag-\enphi\dJ\hrho\,\ephi+O(N^{-1})\end{equation*}}\\ \hline
%%%%%%%%
$\sum_{i=1}^N\hat{\sigma}_-^i\hrho\hat{\sigma}_+^i$
&\parbox{4cm}{\begin{equation*}N \ephi\hrho\,\enphi+O(N^0)\end{equation*}}\\
\hline\hline
%%%%%%%%%%%%%%%%%%%5
$\sum_{i=1}^N\hat{\sigma}_+^i\hrho\hat{\sigma}_+^i$
&\parbox{5cm}{\begin{equation*}\Aop\hrho\Aop-(\Aop)^2\ephi\hrho\enphi+O(N^{-1})\end{equation*}}\\
\hline
%%%%%%%%%%%%%%%%%%%%%%%
$\sum_{i=1}^N\hat{\sigma}_+^i\hrho\hat{s}_z^i$
&\parbox{4cm}{
\begin{align*}\begin{split}
\frac{1}{2}\sqrt{N}\Aop\hrho+O(N^{-1/2})
\end{split}\end{align*}}\\
\hline
%%%%%%%%%%%%%%%%%%%%%%
$\sum_{i=1}^N\hat{\sigma}_-^i\hrho\hat{s}_z^i$
&\parbox{6cm}{
\begin{equation*}
\sqrt{N}\left(\frac{1}{2}\Adag\hrho-\ephi\hrho\enphi\Adag\right)+O(N^{-1/2})
\end{equation*}}\\
\hline
\bottomrule
\end{tabular}
\caption{Replacement rules, to leading non-vanishing order in $1/N$, when $j=1$. The first three lines correspond to typical Lindbladian terms that describe white-noise dephasing (first line), incoherent pumping  (second line) and incoherent decay (third line). The remaining dissipators can be obtained by conjugation of the last three lines.}\label{tab:typLindII}
\end{table*}
\section{Examples}\label{sec:Examples}
In this section we illustrate the full machinery using two simple examples: (i) a collection of atoms undergoing incoherent pumping, decay and white-noise dephasing; (ii) superradiant lasing above the upper threshold. 
\subsection{Pumping+decay+dephasing}
We first study a single particle problem in which an ensemble of two-level atoms with excited state lifetime $\gamma^{-1}$ and inhomogeneous lifetime $2\gamma_d^{-1}$ is incoherently pumped with rate $w$. The master equation describing this evolution is
\begin{equation}
    \partial_t\hrho=\sum_{i=1}^N\big(\gamma\mathcal{D}[\hat{\sigma}_-^i]\hrho+w\mathcal{D}\hat{\sigma}_+^i(\hrho)+\gamma_d\mathcal{D}[\hat{s}_z^i]\hrho\big),
\end{equation}
where $\mathcal{D}[\hat{O}]\hrho=\hat{O}\hrho\hat{O}^\dagger-\{\hat{O}^\dagger\hat{O},\hrho\}/2$ is a standard dissipator. The $\gamma_d$ is the consequence of a white-noise-correlated dephasing process, while the incoherent pumping process results from coherently driving to a rapidly decaying auxiliary level [see Fig.~\ref{fig:Examples:SuperradiantLaser:Schematic}(a)]. This master equation can be solved exactly, leading to the following steady state observables
\begin{equation}\label{eqn:Examples:1:ExpectationValues}
    \begin{matrix}
        \braket{\hat{J}_z}=\frac{N}{2}\left(\frac{w-\gamma}{w+\gamma}\right)&\hspace{0.5cm}\braket{\hat{J}_{x,y}}=0\\[10pt]
        \text{Var}(\hat{J}_z)=\frac{N\gamma w}{(\gamma+w)^2}&\hspace{0.5cm}\text{Var}(\hat{J}_{x,y})=\frac{N}{4}
        \end{matrix}
    \end{equation}
The expressions for the expectation values should be obtained directly from mean field results. The bosonic description will then provide the variances. We begin by calculating the mean field steady state,  which is aligned along $+z$ if $w>\gamma$ and has $j=(w-\gamma)/(w+\gamma)$. Since $j<1$, we use Table~\ref{tab:typLind} to arrive at an effective bosonic description (as promised, the $\sqrt{N}$ contributions cancel among each other)
\begin{align}\begin{split}
    \partial_t\hrho&=(\gamma+w+\gamma_d)\left((\bar{n}+1)\mathcal{D}[\Aop]\hrho+\bar{n}\mathcal{D}[\Adag]\hrho\right)\\[5pt]
    % \Gamma (\bar{n}+1)\mathcal{D}[\Aop]\hrho+\Gamma\bar{n}\mathcal{D}[\Adag]\hrho\\[5pt]
    &+D\big[\hat{q},[\hrho,\hat{q}]\big]+\frac{i(w+\gamma)}{2}\big[\hat{q},\{\hat{l},\hrho\}\big],
\end{split}\end{align}
where $\bar{n}=\gamma(w-\gamma)^{-1}$ and $D=\gamma w/(\gamma+w)$. Thus, the transverse boson is effectively coupled to a bath with decay rate $\gamma+w+\gamma_d$ and thermal occupation $\bar{n}$, while the longitudinal boson $\hat{l}$ undergoes relaxation with rate $(\gamma+w)$ and diffusion with coefficient $D$. The divergence in $\bar{n}$ as $\gamma\to w$ indicates that at $\gamma=w$ the mean field Bloch vector has zero length, and the expansions break down (we are then at the rightmost corner of the Dicke triangle). When $\gamma>w$, the Bloch vector points along $-z$ so it first has to be rotated to $+z$.

From the boson master equation we can immediately calculate transverse variances
\begin{equation}
    \braket{\hat{J}_x^2}\approx\frac{Nj}{4}\braket{(\Aop+\Adag)^2}=\frac{Nj}{4}(2\bar{n}+1)=\frac{N}{4},
\end{equation}
with an identical result for $\braket{\hat{J}_y^2}$, and in agreement with Eq.~(\ref{eqn:Examples:1:ExpectationValues}). For the $\hat{J}_z$ variance, we use $\hat{J}_z=\hat{J}-\Adag\Aop$. To leading order in $1/N$, we then have that $\text{Var}(\hat{J}_z)=\text{Var}(\hat{J})=N\braket{\hat{l}^2}$, which can be calculated directly from the boson description, yielding
\begin{equation}
    N\braket{\hat{l}^2}=N\frac{D}{(\gamma+w)}=\frac{N\gamma w}{(\gamma+w)^2}.
\end{equation}
This is again in agreement with Eq.~(\ref{eqn:Examples:1:ExpectationValues}). 
\subsection{Superradiant laser above upper threshold}
\begin{figure}
    \centering
    \includegraphics[width=0.98\linewidth]{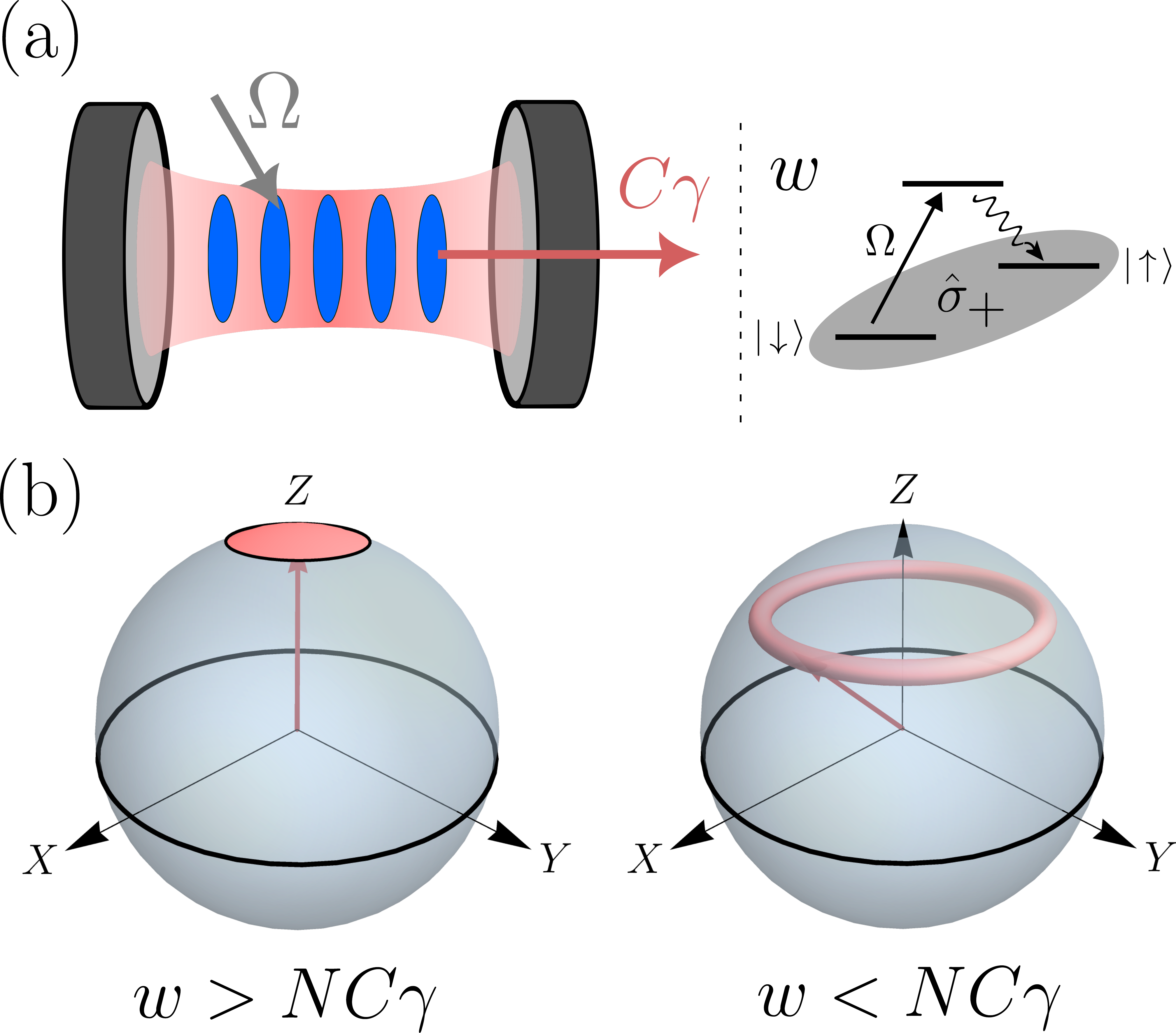}
    \caption{(a) Schematic of a superradiant laser. Atomic excitations are transformed into cavity photons, which then quickly escape the system with rate $C\gamma$ per atom. Population inversion is obtained by coherently driving auxiliary levels. (b) When $w>NC\gamma$ the Bloch vector points along $+z$. When $w<NC\gamma$, the mean field Bloch vector acquires a transverse component ($J_{\text{mf}}^+\neq 0$). At long times, phase diffusion leads to a circular distribution.}
    \label{fig:Examples:SuperradiantLaser:Schematic}
\end{figure}

In this subsection we study a model for superradiant lasing, described by the following master equation~\cite{Meiser2009,Kazakov2013,Zhang2021}
\begin{equation}
    \partial_t\hrho=C\gamma \,\mathcal{D}[\hat{J}^-]+w\sum_{i=1}^N\mathcal{D}[\sigma_-^i]\hrho\equiv\mathcal{L}_{\text{sr}}\hrho,
\end{equation}
which includes incoherent pumping (with rate $w$), and collective emission ($\hat{J}^-$) induced by coupling the atomic transition to a lossy cavity mode [see Fig~\ref{fig:Examples:SuperradiantLaser:Schematic}(a)]. The collective decay rate per particle, $C\gamma$, depends on the cavity cooperativity $C$ and the lifetime of the excited state $\gamma$. We are assuming that $\gamma\ll w,NC\gamma$ so that we can neglect spontaneous emission terms $\mathcal{D}[\hat{\sigma}_-^i]$ in the master equation. Solving the mean field equations under these assumptions leads to two different phases: an incoherent phase when $w>NC\gamma$, with the Bloch vector completely polarized along $+z$ ($J^z_{\text{mf}}=J_{\text{mf}}=N/2$ so $j=1$) and no coherence ($J^{-}_{\text{mf}}=0$); and a lasing phase when $w<NC\gamma$, with nonzero coherence $J_{\text{mf}}^-\neq 0$. 

Fluctuations in the incoherent phase are easier to analyze because the Bloch vector is already pointing along $+z$. Since $j=1$, we need to use the replacement rules in Table~\ref{tab:typLindII}, along with Eq.~(\ref{eqn:DHP:SchwingerBosons}) for collective operators. This leads to
\begin{align}\begin{split}\label{eqn:Examples:SuperradiantLasing:BosonMasterEquation}
    \partial_t\hrho&=w\,\mathcal{D}[\Aop]\hrho+NC\gamma\,\mathcal{D}[\Adag]\hrho\\[5pt]
    &+w\left(\dJ\hrho-\enphi\dJ\hrho \ephi\right)
\end{split}\end{align}
Once again, the transverse boson behaves as if it were connected to a finite temperature bath, now with decay rate $w-NC\gamma$ and thermal occupation $\bar{n}=NC\gamma(w-NC\gamma)^{-1} $, while the dynamics of the longitudinal boson drives the system towards $\delta J=0$ ($J=N/2$). Using the boson description we can directly import known results for steady state two-time correlation functions~\cite{Carmichael_1993,Breuer2007} 
\begin{align}\begin{split}
    \braket{\hat{S}_x(\tau)\hat{S}_x}&\equiv \mathrm{Tr}\left[\hat{S}_xe^{\mathcal{L}_{sr}\tau}\left(\hat{S}_x\hrho_{ss}\right)\right]\\[5pt]
    &\approx\frac{N}{4}\left(\frac{w+NC\gamma}{w-NC\gamma}\right)e^{-(w-NC\gamma)\tau}
\end{split}\end{align}
and then obtain power spectral densities. As $w\to NC\gamma$, fluctuations (encoded in $\bar{n}$) diverge, indicating that a more careful analysis (deferred to Sec.~\ref{sec:PhaseTransitions}) is required. The decay constant $w-NC\gamma$ could be identified as the ``linewidth" of the emitted light in the incoherent phase, and is in agreement with Eq.~(4) of Ref.~\cite{Debnath2018} in the appropriate limit (strong pumping, no dephasing, negligible spontaneous emission and large cavity decay rate). 

When $w<NC\gamma$, the mean field Bloch vector no longer points along $+z$, and its transverse direction can be arbitrarily chosen to lie along the $+x$ direction due to the weak $U(1)$ phase symmetry of the system. To study fluctuations, the Bloch vector first needs to be rotated onto the $+z$ axis, and only then should the replacement rules be used. In this configuration, the $\hat{S}_y$ variable is a proxy for the azimuthal phase of the Bloch vector, will undergo diffusion, and its two-time correlation function will encode the linewidth of the laser (see Appendix~\ref{app:SuperradiantLaserAboveThreshold}). Because of the diffusive behaviour phase fluctuations will grow with time, at long times the large $N$ expansion about the mean field steady state will break down, and the distribution will become symmetrical about rotations along the $z$ axis, as depicted in Fig.~\ref{fig:Examples:SuperradiantLaser:Schematic}(b), although modified large $N$ techniques are still applicable.

All of the results presented so far can also be obtained using second-order cumulant techniques, although we believe that the boson formalism provides a simple picture of the longitudinal fluctuations. However, paralleling the purely Hamiltonian case, the formalism can be extended to describe the vicinity of the phase transition in a controlled way.

\section{Driven-dissipative phase transitions}\label{sec:PhaseTransitions}
In this section we address item $(\Pi)$ in the introduction and show how the bosonic representation can be used to describe the properties at, and in the vicinity of, driven-dissipative phase transitions in all-to-all models. We will consider two examples: the first one will be a continuation of the analysis of the superradiant laser in Section~\ref{sec:Examples} and the second one will be a dissipative generalization of the all-to-all transverse field Ising model, previously analyzed in Ref.~\cite{Paz2021}. 
\subsection{Superradiant laser near upper threshold}\label{subsec:DDPhaseTransitions:SuperradiantLaser}
Here we describe the onset of the lasing transition when $N\to\infty$. Equation~(\ref{eqn:Examples:SuperradiantLasing:BosonMasterEquation}) naively indicates that fluctuations diverge as $w\to NC\gamma\equiv w_c$. As in the Hamiltonian case, this is a breakdown of the large $N$ approximation as implemented by Table~\ref{tab:typLindII}. We expect instead that higher order terms in the $1/N$ expansion will stabilize the system. To pursue this, we need to extend Table~\ref{tab:typLindII} by including further corrections. We show the resulting replacement rules in Table~\ref{tab:DDPhaseTransitions:NLOReplacementRules}. For completeness, we also include the expansion of collective operators $\hat{J}_{\pm,z}$.

\begin{table*}[h!]
\centering
\setlength{\tabcolsep}{10pt}
\begin{tabular}{c||c}\toprule 
 Spin term &Boson term\\[5pt] \hline\hline
\parbox{4cm}{
\begin{equation*}
\mathcal{D}_{zz}(\hrho)\equiv \sum_{i=1}^N\hat{s}_z^i\hrho\hat{s}_z^i
\end{equation*}}
%%%%%%%%
&\parbox{12cm}{
\begin{align*}\begin{split}
&\frac{N\hrho}{4} +\left(\Aop \ephi\hrho \,\enphi\Adag-\frac{1}{2}\{\Adag\Aop,\hrho\}\right)\\[5pt]
-&{\color{darkgreen}\frac{1}{N}\bigg[\Aop\ephi\dJ\hrho\enphi\Adag+\Adag\enphi\dJ\hrho\ephi\Aop+\frac{1}{2}\Aop\{\Adag\Aop,\ephi\hrho\enphi\}\Adag}\bigg]\\[5pt]
+&{\color{darkgreen}\frac{1}{N}\left[\dJ\hrho+\dJ\{\Adag\Aop,\hrho\}+\Adag\Aop\hrho\Adag\Aop\right]}+O(N^{-2})
\end{split}\end{align*}
}\\ \hline
%%%%%%%%
\parbox{4cm}{\begin{equation*}\mathcal{D}_{+-}(\hrho)\equiv \sum_{i=1}^N\hat{\sigma}_+^i\hrho\hat{\sigma}_-^i\end{equation*}}
&\parbox{12cm}{\begin{align*}\begin{split}
&\left[\Aop\hat{\rho}\Adag-\enphi\dJ\hrho\,\ephi\right]{\color{darkgreen}-\frac{1}{N}\left[2\dJ\Aop\hrho\Adag+\frac{1}{2}\{\Adag\Aop,\Aop\hrho\Adag\}\right]}\\[5pt]
+&{\color{darkgreen}\frac{1}{N}\bigg[\{\Adag\Aop,\enphi\dJ\hrho\ephi\}+(\Aop)^2\ephi\hrho\enphi(\Adag)^2\bigg]}+O(N^{-2})
\end{split}\end{align*}}\\ \hline
%%%%%%%%
\parbox{4cm}{\begin{equation*}\mathcal{D}_{-+}(\hrho)\equiv \sum_{i=1}^N\hat{\sigma}_-^i\hrho\hat{\sigma}_+^i\end{equation*}}
&\parbox{12cm}{\begin{align*}\begin{split}
&N \ephi\hrho\,\enphi+{\color{darkgreen}\left[\dJ \ephi\hrho\enphi+\Adag\hrho\Aop-\{\Adag\Aop,\ephi\hrho\enphi\}\right]}+O(N^{-1})
\end{split}\end{align*}}\\
\hline\hline

%%%%%%%%%%%%%%%%%%%5
\parbox{4cm}{\begin{equation*}\mathcal{D}_{++}(\hrho)\equiv \sum_{i=1}^N\hat{\sigma}_+^i\hrho\hat{\sigma}_+^i\end{equation*}}
&\parbox{5cm}{\begin{equation*}\Aop\hrho\Aop-(\Aop)^2\ephi\hrho\enphi+O(N^{-1})\end{equation*}}\\
\hline
%%%%%%%%%%%%%%%%%%%%%%%%%%%%%%%%%%%%%%%%%
%%%%%%%%%%%%%%%%%%%%%%%%%%%%%%%%%%%%%%%%%
\parbox{4cm}{
\begin{equation*}
\mathcal{D}_{+z}(\hrho)\equiv\sum_{i=1}^N\hat{\sigma}_+^i\hrho\hat{s}_z^i
\end{equation*}}

&\parbox{12cm}{
\begin{align*}\begin{split}
&\frac{\sqrt{N}\Aop\hrho}{2}{\color{darkgreen}+\frac{1}{\sqrt{N}}\bigg[\enphi\hrho\dJ\ephi\Aop+(\Aop)^2\ephi\hrho\Adag\bigg]}\\[5pt]
-&{\color{darkgreen}\frac{1}{\sqrt{N}}\bigg[\frac{\dJ\Aop\hrho}{2}+\frac{\Adag(\Aop)^2\hrho}{4}+\Aop\hrho\Adag\Aop\bigg]}+O(N^{-3/2})
\end{split}\end{align*}}\\
\hline

\parbox{4cm}{
\begin{equation*}
 \mathcal{D}_{-z}(\hrho)\equiv\sum_{i=1}^N\hat{\sigma}_-^i\hrho\hat{s}_z^i
\end{equation*}}

&\parbox{12cm}{
\begin{align*}\begin{split}
&\sqrt{N}\left(\frac{\Adag\hrho}{2}-\ephi\hrho\enphi\Adag\right){\color{darkgreen}-\frac{1}{2\sqrt{N}}\bigg[\dJ\Adag\hrho+(\Adag)^2\Aop\hrho+2\Adag\hrho\Adag\Aop\bigg]}\\[5pt]
+&{\color{darkgreen}\frac{1}{\sqrt{N}}\bigg[\Adag\Aop\ephi\hrho\enphi\Adag+\frac{1}{2}\ephi\hrho\enphi\Adag\Aop\Adag+\frac{1}{2}\ephi\hrho\enphi\Adag\bigg]}+O(N^{-3/2})
\end{split}\end{align*}}\\
\hline
\hline
\parbox{4cm}{
\begin{equation*}
    \hat{J}_+
\end{equation*}
}
&\parbox{12cm}{
\begin{align*}
    \begin{split}
         \sqrt{N}\Aop+{\color{darkgreen}\frac{1}{N}\dJ\Aop-\frac{1}{2N}\Adag\Aop\Aop}
    \end{split}
\end{align*}}\\
\hline
\parbox{4cm}{
\begin{equation*}
    \hat{J}_z
\end{equation*}
}
&\parbox{12cm}{
\begin{align*}
    \begin{split}
        \frac{N}{2}+ \dJ-\Adag\Aop
    \end{split}
\end{align*}}\\

\hline
\bottomrule
\end{tabular}
\caption{Lindbladian terms to higher order in $1/N$ (highlighted in green). All other combinations can be obtained by complex conjugation. We also include expressions for collective spin operators}\label{tab:DDPhaseTransitions:NLOReplacementRules}
\end{table*}

Setting $w=w_c+\delta w$ in Eq.~(\ref{eqn:Examples:SuperradiantLasing:BosonMasterEquation}), with the understanding that $\delta w\ll NC\gamma$, leads to the following bosonic master equation
\begin{widetext}
\begin{align}\label{eqn:DDPhasTransition:SuperradiantMasterEquation}
    \begin{split}
        \partial_t\hrho&=(w_c+\delta w)\mathcal{D}[\Aop]\hrho+w_c\left(\dJ\hrho-\enphi\dJ\hrho\,\ephi\right)+\frac{w_c}{N}\mathcal{D}[\Aop \ephi](\Aop\hrho\Adag)-\frac{2w_c}{N}\enphi\left\{\mathcal{D}[\Aop\ephi](\dJ\hrho)\right\}\ephi\\[10pt] 
        &+w_c\,\mathcal{D}[\Adag]\hrho+\frac{2w_c}{N}\mathcal{D}[\Adag](\dJ\hrho)+\frac{w_c}{2N}\left(\{\Aop\Adag\Adag\Aop,\hrho\}-\Adag\{\Adag\Aop,\hrho\}\Aop\right)
    \end{split}
\end{align}
\end{widetext}
We will now begin with the simplifications. First, even though at the critical point the transverse boson occupation diverged and timescales associated to it became very long $(w-NC\gamma)^{-1}$, the longitudinal boson was still forced to decay to $\delta J=0$ within a finite timescale $(NC\gamma)^{-1}$. Thus we can project the system into the steady state subspace of the longitudinal boson. Mathematically, this is performed by computing the matrix element of the evolution superoperator (using the trace inner product) between the right/left steady states of the longitudinal boson, which are $|\delta J=0\rangle\hspace{-0.05cm}\langle\delta J=0|$ and the identity, respectively. This yields an equation for the reduced density matrix of the transverse boson $\hrho_T$:
\begin{align}\label{eqn:DDPhasTransition:SuperradiantMasterEquation2}
    \begin{split}
        \partial_t\hrhoT&=(w_c+\delta w)\mathcal{D}[\Aop]\hrhoT+w_c\,\mathcal{D}[\Adag]\hrhoT\\[5pt]
        &+\frac{w_c}{N}\mathcal{D}[\Aop](\Aop\hrhoT\Adag)\\[5pt]
        &+\frac{w_c}{2N}\left(\left[\Aop,(\Adag)^2\Aop\hrhoT\right]+\left[\hrhoT\Adag(\Aop)^2,\Adag\right]\right),
    \end{split}
\end{align}
which ought to describe the critical properties of the lasing transition. 

Given that $\bar{n}$ diverges in the linear theory and that it is stabilized by $1/N$ corrections, we expect that the size of the steady state distribution in boson phase space will scale with $N$ and hence be very large, much larger than the size of quantum noise. This implies that we can treat $\Aop,\Adag $, when suitably normalized, as classical variables. To take the classical limit, we introduce $\hat{\alpha}=N^{-f_A}\Aop$, where $f_A$ is a number to be determined, with commutation relations $[\hat{\alpha}^{\vpd},\hat{\alpha}^\dagger]=N^{-2f_A}$. Since we want the $\hat{\alpha}$ variables to be of size $\sim 1$ in the classical limit, we need to consider an effective Planck constant equal to $N^{-2f_A}$. Thus, when $N\to\infty$ commutators become Poisson brackets $[\,\,,\,]\approx iN^{-2f_A}\{\,\,,\,\}^{\text{pb}}$. This leads to $\{\alpha,\bar{\alpha}\}^{\text{pb}}=-i$, where we are now treating $\alpha,\bar{\alpha}$ as classical commuting variables. To take the classical limit of Eq.~(\ref{eqn:DDPhasTransition:SuperradiantMasterEquation2}), we express it as much as possible in terms of commutators (and sometimes double commutators). 
% For instance,
% \begin{align}\begin{split}
%     2\mathcal{D}[\hat{O}]\hrho&=[\hat{O}\hrho,\hat{O}^\dagger]+[\hat{O},\hrho\hat{O}^\dagger]\\
%     &\to N^{-2f_A}i\left(\{O,\rho_c\bar{O}\}^{\text{pb}}+\{O\rho_c,\bar{O}\}\right),
% \end{split}\end{align}
% where $O,\bar{O},\rho_c$, which are functions of $\alpha$ and $\bar{\alpha}$, are the classical analogues of $\hat{O},\hat{O}^\dagger,\hrho$. Another important combination is $2\mathcal{D}[\hat{O}]\hrho+2\mathcal{D}[\hat{O}^\dagger]\hrho=[\hat{O},[\hrho,\hat{O}^\dagger]]+[\hat{O}^\dagger,[\hrho,\hat{O}]]$, which has a double commutator structure, and hence becomes $-N^{-4f_A}\{O,\{\rho,\bar{O}\}^{\text{pb}}\}^{\text{pb}}-N^{-4f_A}\{\bar{O},\{\rho,O\}^{\text{pb}}\}^{\text{pb}}$. 
The resulting classical master equation for $\rho_c$ (the classical analogue of $\hat{\rho}$) is
\begin{align}\label{eqn:DDPhasTransition:SuperradiantMasterEquation3}
    \begin{split}
        \partial_t\rho_c&=\frac{i\delta w}{2}\left(\{\alpha,\rho_c\bar{\alpha}\}^{\text{pb}}+\{\alpha\rho_c,\bar{\alpha}\}^{\text{pb}}\right)\\
        &-\frac{w_c}{N^{2f_A}}\{\alpha,\{\rho_c,\bar{\alpha}\}^{\text{pb}}\}^{\text{pb}}\\
        &+\frac{iw_c}{N^{1-2f_A}}\left(\bar{\alpha}\{\alpha^2\rho_c,\bar{\alpha}\}^{\text{pb}}+\alpha\{\alpha,\rho_c\bar{\alpha}^2\}^{\text{pb}}\right)
    \end{split}
\end{align}
In the previous equation, the first two lines came from the linear theory, while the third line is the nonlinearity. We can make all the terms of the same size if we choose $f_A=1/4$ and keep $\zeta=N^{1/2}\delta w/w_0$ fixed as $N\to \infty$. Then the partial differential equation governing the evolution of $\rho_c$, the classical probability distribution, is
\begin{align}\label{eqn:DDPhasTransition:SuperradiantMasterEquation4}
    \begin{split}
        \frac{\partial\rho_c}{\partial(w_0t/\sqrt{N})}&=\frac{\zeta}{2}\left[\partial_{\bar{\alpha}}(\bar{\alpha}\rho_c)+\partial_{\alpha}(\alpha \rho_c)\right]+\partial_{\alpha\bar{\alpha}}^2\rho_c\\
        &+\alpha\partial_{\bar{\alpha}}(\bar{\alpha}^2\rho_c)+\bar{\alpha}\partial_{\alpha}(\alpha^2\rho_c)
    \end{split}
\end{align}
This equation determines the cross-over behaviour in the vicinity of the phase transition point ($\zeta=0$), manifestly shows that timescales are slowed down by a factor of $\sqrt{N}$ and demonstrates that the system is self-similar as $N$ increases, provided $\zeta$ is kept fixed. The steady state of the equation can in fact be written down analytically,
\begin{equation}
    \rho_c^{\text{ss}}=\mathcal{N}(\zeta) \exp\left(-|\alpha|^4-\zeta |\alpha|^2\right),
\end{equation}
where $\mathcal{N}(\zeta)$ is a normalization factor with respect to the measure $d\alpha d\bar{\alpha}$. When $\zeta>0$ (incoherent regime), the distribution is peaked at 0 and becomes gaussian when $\zeta\gg1$. Similarly, when $\zeta\ll -1$, the distribution is also gaussian, but now peaked at $|\alpha|^2=-\zeta/2$, indicating a nonzero value of $\braket{\hat{J}^+\hat{J}^-}$. In the vicinity of $\zeta=0$, the distribution is non-gaussian.

\begin{figure*}
    \centering
    \includegraphics[width=0.92\linewidth]{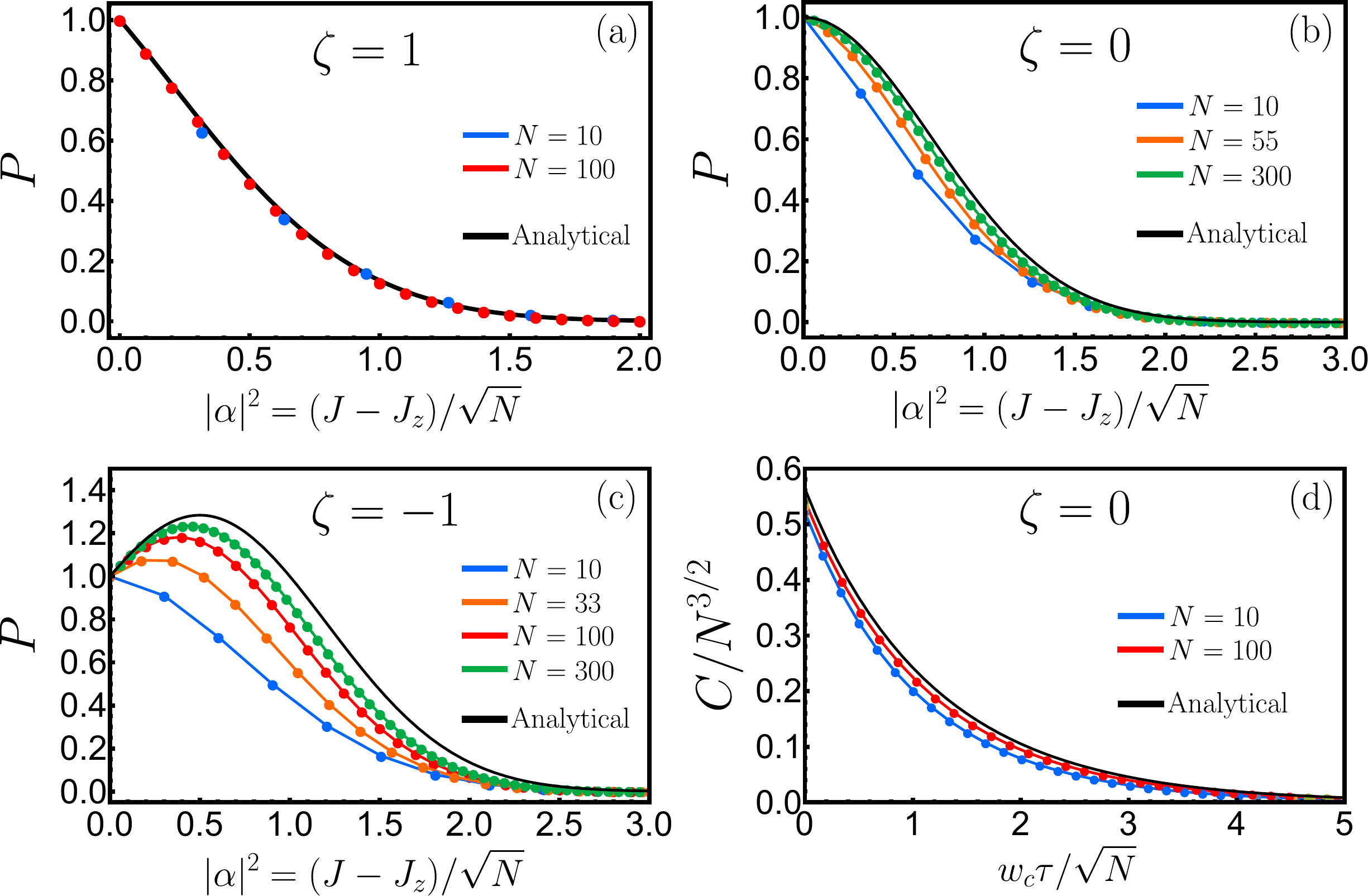}
    \caption{(a) Probability distribution $P=\rho_c^{ss}/\mathcal{N}(\zeta)=e^{-|\alpha|^4-\zeta|\alpha|^2}$, normalized such that $P=1$ at $\alpha=0$, as a function of $|\alpha|^2$ for $\zeta=1$ (incoherent phase) and $N=10,100$. (b) Same as (a), but for $\zeta=0$ (transition point) and $N=10,55,300$. (c) Same as (a) , but for $\zeta=-1$ (coherent phase) and $N=10,33,100,300$. (d) First order coherence function as a function of time for $N=10,100$ versus analytical profile obtained from Eq.~(\ref{eqn:DDPhaseTransition:Superradiance:FirstCoherence}).}
    \label{fig:DDPhaseTransition:Superradiance}
\end{figure*}

We benchmark this analytical result against numerical simulation of Eq.~(\ref{eqn:DDPhasTransition:SuperradiantMasterEquation}) for $\zeta=-1,0,1$ and different values of $N$ up to $N=300$. To obtain $\rho_c^{ss}$ from these numerical results, we identify $|\alpha|^2$ with $\Adag\Aop/\sqrt{N}=(\hat{J}-\hat{J}_z)/\sqrt{N}$ and plot the resulting probabilities in Fig.~\ref{fig:DDPhaseTransition:Superradiance} (a), (b) and (c) against the analytical formula. We find good agreement when the distribution is peaked around $|\alpha|^2=0$, especially at larger $N$, and observe a trend towards convergence as $N$ increases when the peak is at finite $|\alpha|^2$. In general, we expect corrections to be of relative size $N^{-1/2}$ rather than $N^{-1}$ due to the scalings around the critical point, indicating that even at $N=300$ we can expect errors of about $6\%$.

The classical master equation Eq.~(\ref{eqn:DDPhasTransition:SuperradiantMasterEquation4}) defines a classical generator of time evolution $\mathcal{M}_{\text{sr}}$, which encodes more information than just the steady state distribution $\rho_c^{ss}$. In particular, correlation functions of the quantum system can be calculated, to leading order in $1/N$, using $\mathcal{M}_{\text{sr}}$. For example, the two-point function 
\begin{align}\begin{split}\label{eqn:DDPhaseTransition:Superradiance:FirstCoherence}
    C(\tau)&=\mathrm{Tr}\left[\hat{J}^+ e^{\mathcal{L}_{\text{sr}}\tau}(\hat{J}^-\hat{\rho}_{\text{ss}})\right]\\
    &\approx N^{3/2}\int \alpha e^{\mathcal{M}_{\text{sr}}\tau}(\bar{\alpha}\rho_c^{\text{ss}})d\alpha d\bar{\alpha}
\end{split}\end{align}
can be expressed entirely in terms of the classical master equation, with a predicted scaling of $N^{3/2}$. This is illustrated in Fig.~\ref{fig:DDPhaseTransition:Superradiance}(d), which verifies the $N^{3/2}$ scaling of $C(\tau)$ and demonstrates good agreement between the full numerical solution of Eq.~(\ref{eqn:DDPhasTransition:SuperradiantMasterEquation}) and the classical formulas.
\subsection{Dissipative all-to-all transverse field Ising model}\label{subsec:dissipativeTFIM}
Here we consider a dissipative version of the transverse field Ising model, defined by the following master equation
\begin{equation}\label{eqn:DDPhaseTransitions:TFIMMasterEquation0}
    \partial_t\hrho=-i\left[-\Delta \hat{J}_z-\frac{g}{N}\hat{J}_x^2,\hrho\right]+\gamma\sum_{i=1}^N\mathcal{D}[\hat{\sigma}_+^i]\hrho\equiv\mathcal{L}_{\text{tf}}\hrho,
\end{equation}
which is a generalization of the Hamiltonian model studied in Sec.~\ref{subsec:Motivation} that includes incoherent pumping from the $\ket{\downarrow}\to\ket{\uparrow}$ states (this is equivalent to the model studied in Ref.~\cite{Paz2021} after a rotation by $\pi$ about the $x$ axis). Assuming $\Delta>0$, the system displays two mean field phases: a disordered phase, with $J_z^{\text{mf}}=N/2$ and $J_{x,y}^{\text{mf}}=0$, and an ordered phase, with $J_{x,y,z}^{\text{mf}}\neq 0$, and a phase boundary defined by $\gamma=2\sqrt{\Delta(g-\Delta)}$. We will approach the critical boundary from the disordered phase because the Bloch vector is already aligned along $+z$, and we will do so by varying $\Delta$ while keeping $g$ fixed. Since $j=1$ in this phase, we use the replacement rules from Table~\ref{tab:typLindII}, which to leading order give rise to the following bosonic master equation:
\begin{align}\label{eqn:DDPhaseTransitions:TFIMMasterEquation1}
    \begin{split}
        \partial_t\hrho&=-\frac{i}{2}\left[\Delta\hat{p}^2+(\Delta-g)\hat{x}^2,\hrho\right]+\gamma\mathcal{D}[\Aop]\hrho\\
        &+i\Delta\left[\dJ,\hrho\right]+\gamma\left(\dJ\hrho-\enphi\dJ\ephi\,\right),
    \end{split}
\end{align}
where the quadratures $\hat{x}=(\Aop+\Adag)/\sqrt{2}$ and $\hat{p}=-i(\Aop-\Adag)/\sqrt{2}$ are defined as before. Once again, the longitudinal boson evolution just drives the system to the state $\ket{\delta J=0}$. The instability towards the ordered phase as $\Delta$ is reduced can then be interpreted in the boson language as being caused by the switch from a regular to an inverted parabolic potential when $\Delta<g$. Dissipation provides some stabilization, reducing the range of $\Delta$ for which there is an ordered phase, but if $\gamma$ is small there will still be an instability as $\Delta$ is further reduced. If $\gamma$ is large enough, the instability disappears and the disordered phase is always stable. 

We will work in a regime of vanishingly small dissipation, with $\gamma$ scaling with $N$ in a yet-to-be-determined way, and leave the analysis of finite $\gamma$ to Appendix~\ref{app:FiniteGamma}. As in Sec.~\ref{subsec:Motivation}, in the vicinity of the phase transition the system will be stabilized by a quartic nonlinearity coming from the Hamiltonian. Guided from our experience in Sec.~\ref{subsec:Motivation}, we expect fluctuations in $\hat{x}$ and $\hat{p}$ to behave differently, so we express the master equation, now including the non-linearity, in terms of the quadratures. We also project out the longitudinal boson and work with the reduced density matrix for the transverse boson $\hrhoT$
\begin{align}\label{eqn:DDPhaseTransitions:TFIMMasterEquation2}
    \begin{split}
        \partial_t\hrho&=-\frac{i}{2}\left[\Delta\hat{p}^2+(\Delta-g)\hat{x}^2+\frac{g\hat{x}^4}{2N},\hrhoT\right]\\
        &+\frac{\gamma}{2}\mathcal{D}[\hat{x}]\hrhoT+\frac{i\gamma}{4}\left(\left[\{\hat{p},\hrhoT\},\hat{x}\right]+\left[\hat{p},\{\hat{x},\hrhoT\}\right]\right).
    \end{split}
\end{align}
Note that we have omitted a $\mathcal{D}[\hat{p}]\hrho_T$ term because fluctuations in $\hat{p}$ will be smaller than fluctuations in $\hat{x}$. It turns out that, unlike the case of the Hamiltonian model of Sec.~\ref{subsec:Motivation}, fluctuations in $\hat{p}$ will not be reduced, but will instead stay of the same size without any $N$ dependence. Given that fluctuations in $\hat{x}$ are still enhanced, the distribution in phase space will be large compared to the size of the quantum noise, and we can treat it as a classical probability distribution. Thus, if we introduce a scaled $\hat{z}=\hat{x}N^{-f_x}$, which satisfies $[\hat{z},\hat{p}]=iN^{-f_x}$, we should replace commutators by Poisson brackets according to $[\,,\,]\approx iN^{-f_x}\{\,,\,\}^{\text{pb}}$. We then simultaneously scale the distance to the critical point by defining $\eta$ using $\Delta=g+\eta \Delta N^{-f_{\Delta}}$, scale the strength of dissipation by introducing a reduced $\gamma_{\text{red}}=\gamma N^{-f_\gamma}$, and demand that all the terms be of the same size in Eq.~(\ref{eqn:DDPhaseTransitions:TFIMMasterEquation2}). This leads to $f_x=1/4$, $f_\Delta=1/2$ and $f_\gamma=1/4$, and to a classical master equation for the probability distribution $\rho_c$, which is a function of $z$ and $p$ (i.e. the classical analogues of $\hat{z}$ and $\hat{p}$),
\begin{align}\begin{split}\label{eqn:DDPhaseTransitions:TFIMMasterEquation3}
    \frac{\partial\rho_c}{\partial(\Delta t/N^{1/4})}&=\big[(\eta z+z^3)\partial_p\rho_c-p\partial_z\rho_c\big]\\
    &+\frac{\gamma_{\text{red}}}{4\Delta}\left[\partial_p^2\rho_c+2\partial_p(p\rho_c)+2\partial_z(z\rho_c)\right].
\end{split}\end{align}
The first line of Eq.~(\ref{eqn:DDPhaseTransitions:TFIMMasterEquation3}) is just Hamiltonian flow, with a classical Hamiltonian $H_c=(2p^2+2\eta z^2+z^4)/4$, while the second line introduces diffusion and relaxation. The steady state solution can be written down analytically
\begin{align}\begin{split}\label{eqn:DDPhaseTransitions:TFIMSolutionSteadyState}
    \rho_c^{ss}&=\mathcal{N}\exp\bigg[-2\left(p-\frac{\gamma_{\text{red}}z}{2\Delta}\right)^2\\
    &\hspace{3.2cm}-2\left(\eta+\frac{\gamma_{\text{red}}^2}{4\Delta^2}\right)z^2-z^4\bigg],
\end{split}\end{align}
is Boltzmann like, and reduces to $\exp(-4H_c)$ in the limit $\gamma_{\text{red}}\ll\Delta$ ($\mathcal{N}$ is a normalization factor). 

\begin{figure*}
    \centering
    \includegraphics[width=0.92\linewidth]{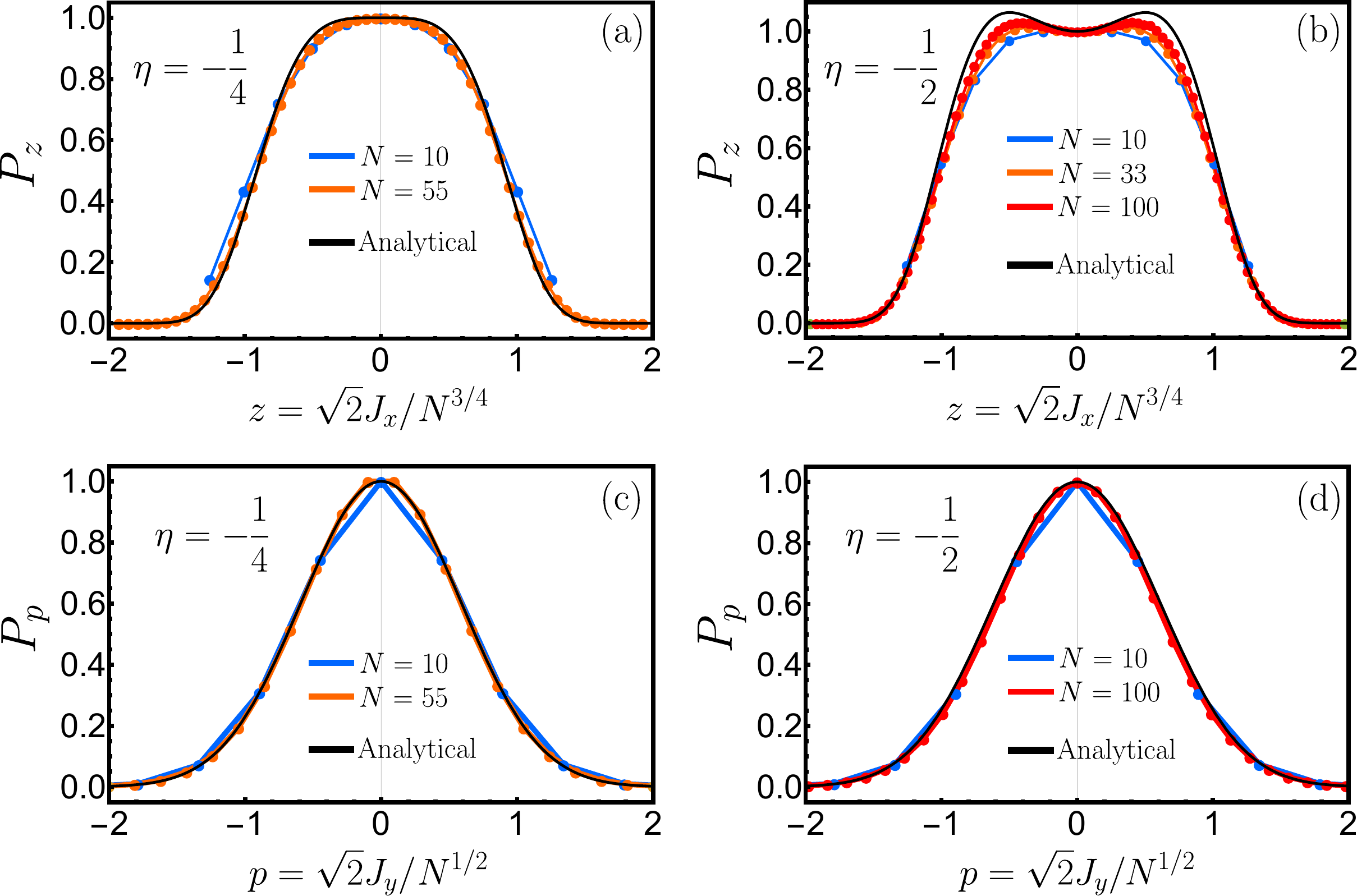}
    \caption{(a) Marginal probability distribution $P_z$, obtained by integrating Eq.~(\ref{eqn:DDPhaseTransitions:TFIMSolutionSteadyState}) over $p$ and fixing $P_z=1$ at $z=0$, for $N=10,55$ and $\eta=-1/4$. (b) Same as (a) but with $N=10,33,100$ and $\eta=-1/2$. (c) Marginal probability distribution $P_p$, obtained by integrating Eq.~(\ref{eqn:DDPhaseTransitions:TFIMSolutionSteadyState}) over $z$ and fixing $P_p=1$ at $p=0$, for $N=10,55$ and $\eta=-1/4$. (d) Same as (c) but with $N=10,100$ and $\eta=-1/2$.}
    \label{fig:DDPhaseTransitionsTFIM}
\end{figure*}

To benchmark this solution, we first make contact with the original spin variables by recalling that, to leading order, $\hat{J}_x=\hat{x}\sqrt{N/2}=\hat{z}N^{3/4}/\sqrt{2}$ and $\hat{J}_y=\hat{p}\sqrt{N/2}$. We then simulate Eq.~(\ref{eqn:DDPhaseTransitions:TFIMMasterEquation0}) to obtain the steady state for $N$ up to $100$, calculate the probability distributions associated to the $\hat{J}_{x},\hat{J}_y$ operators, and compare them against the marginal probability distributions obtained from Eq.~(\ref{eqn:DDPhaseTransitions:TFIMSolutionSteadyState}). The results are shown in Fig.~\ref{fig:DDPhaseTransitionsTFIM} for $\gamma_{\text{red}}/\Delta=1$ and $\eta=-1/2,-1/4$, demonstrating good agreement for all of them. Note that the critical $e^{-z^4}$ profile is obtained at $\eta=-\gamma_{\text{red}}^2/(4\Delta^2)$ rather than $\eta=0$. This can also be derived using the equation for the mean field critical boundary after replacing $\Delta,\gamma$ in favor of $\eta,\gamma_{\text{red}}$ and letting $N\to\infty$.

\begin{figure}
    \centering
    \includegraphics[width=0.98\linewidth]{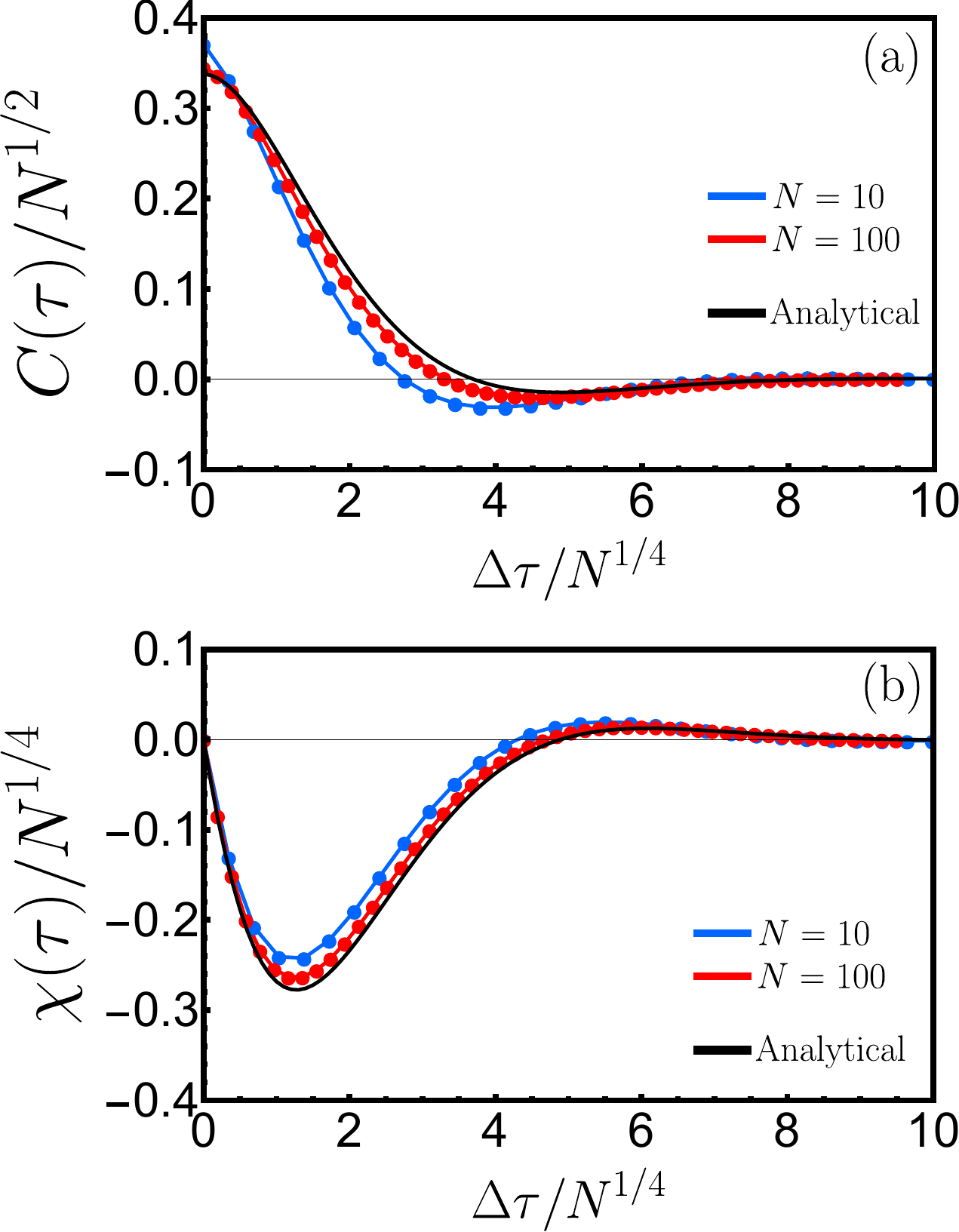}
    \caption{(a) Scaled correlation function $C(\tau)/N^{1/2}$ as a function of scaled time $\Delta \tau/N^{1/4}$ for $N=10,100$, $\gamma_{\text{red}}=\Delta$ and $\eta=-1/4$. (b) Scaled response function $\chi(\tau)/N^{1/4}$ as a function of scaled time $\Delta \tau/N^{1/4}$ for $N=10,100$, $\gamma_{\text{red}}=\Delta$ and $\eta=-1/4$.}
    \label{fig:DDPhaseTransitionsTFIM2}
\end{figure}

As in the case for the superradiant transition, Eq.~(\ref{eqn:DDPhaseTransitions:TFIMMasterEquation3}) defines a classical generator of time evolution $\mathcal{M}_{\text{tf}}$ that encodes the response of the system to perturbations. We can use it to calculate correlation and response functions such as
\begin{align}
    \begin{split}
        C(\tau)&=\frac{\braket{\{\hat{J}_x(\tau),\hat{J}_x(0)\}}}{N}\approx N^{1/2}\,\overline{z(\tau)z(0)}\\
        \chi(\tau)&=\frac{\braket{[\hat{J}_x(\tau),\hat{J}_x(0)]}}{iN}\approx \frac{N^{1/4}}{2}\overline{\{z(\tau),z(0)\}^{\text{pb}}},
    \end{split}
\end{align}
where $\hat{J}_x(\tau)=e^{\mathcal{L}^{\ddagger}_{\text{tf}}\tau}(\hat{J}_x)$, $\mathcal{L}^{\ddagger}_{\text{tf}}$ is the adjoint of $\mathcal{L}_{\text{tf}}$ with respect to the trace inner product, the overlines are averages with respect to $\rho_c^{ss}$ in Eq.~(\ref{eqn:DDPhaseTransitions:TFIMSolutionSteadyState}), $z(\tau)=e^{\mathcal{M}_{\text{tf}}^\ddagger\tau}(z)$, and $\mathcal{M}^\ddagger_{\text{tf}}$ is the adjoint of $\mathcal{M}_{\text{tf}}$ with respect to the inner product on phase space (integration over $z$ and $p$). After some manipulations, we get
\begin{align}
    \begin{split}
        C(\tau)&\approx N^{1/2}\int dz\,dp \,z\,e^{\mathcal{M}_{\text{tf}}\tau}(z\rho_c^{ss})\\
        \chi(\tau)&\approx \frac{N^{1/4}}{2}\int dz\,dp \,z\,e^{\mathcal{M}_{\text{tf}}\tau}(\partial_p\rho_c^{ss}).
    \end{split}
\end{align}
These scalings are consistent with the scalings reported in Ref.~\cite{Paz2021}. We compare these formulas against numerical solution of Eq.~(\ref{eqn:DDPhaseTransitions:TFIMMasterEquation0}) in Fig.~(\ref{fig:DDPhaseTransitionsTFIM2}), demonstrating good agreement that improves as $N$ increases.

\section{Thermal behaviour}\label{sec:Thermal}
In this section, we show that the generalized boson mappings can also be used to study thermal behaviour of collective Hamiltonians. Consider the Dicke model~\cite{Kirton2019}, which hosts both ground state~\cite{Emary2003} and thermal phase transitions~\cite{Hepp1973,Wang1973}, and is defined by the Hamiltonian
\begin{equation}
    \hat{H}_D=\omega \hat{c}^\dagger\hat{c}-\omega_0\hat{J}_z+\frac{2\lambda}{\sqrt{N}}(\hat{c}+\hat{c}^\dagger)\hat{J}_x,
\end{equation}
where $\hat{c},\hat{c}^\dagger$ describe a bosonic mode (often photons in a cavity~\cite{Baumann2010} or motional modes in an ion crystal~\cite{Safavi2018}). The thermal properties of the system are encoded in the partition function
\begin{equation}\label{eqn:ThermalPhysics:Z}
    Z_D=\mathrm{Tr}(e^{-\beta\hat{H}_D}),
\end{equation}
where $\beta=1/T$ is the inverse temperature. To be more precise, the trace is taken over the $2^N$ dimensional Hilbert space of $N$ independent spin $1/2$, not only over the permutationally symmetric Dicke states~\cite{Alcalde2012,PerezFernandez2017}. Mathematically, this partition function is often calculated (in the large $N$ limit) by representing the trace as an integral and performing a saddle point approximation~\cite{Wang1973}. However, our treatment will be closer in spirit to the purely Hamiltonian analysis of the Tavis-Cummings model of Ref.~\cite{HEPP1973360}.

The thermal state $e^{-\beta\hat{H}_D}$ is weakly permutationally symmetric, so it is amenable to analysis by means of the operator mappings of Sec.~\ref{sec:DissipativeMapping}. Since only collective spin operators appear, we will only need the lower entries of Table~\ref{tab:DDPhaseTransitions:NLOReplacementRules}. The only subtlety is that the bosonic representation only counts each different spin length sector $J$ once, so we need to take into account explicitly the degeneracy of each different $J$, given by
\begin{equation}
    d_J=\frac{N!(2J+1)}{(N/2-J)!(N/2+J+1)!}.
\end{equation}
This entropic factor, when balanced against energetics, will end up determining the average spin length $Nj/2$. To take this into account, we define an effective Hamiltonian
\begin{equation}
    \hat{K}=\beta\hat{H}_D-\log(d_{\hat{J}}),
\end{equation}
where we have also included the dependence with $\beta$ because it is a tunable parameter. The thermal state is now $e^{-\hat{K}}$, and the explicit inclusion of $d_J$ means that we now have to use the bosonic representation of $\hat{H}_D$.

As in previous examples, we first examine the mean field behaviour. Since the spin length is expected to be $j<1$, we use the type I replacement $\hat{J}=Nj/2+N^{1/2}\hat{l}$ to express the degeneracy factor as
\begin{align}\begin{split}
\log(d_{\hat{J}})=-\frac{N}{2}f(j)-f'(j)N^{1/2}\hat{l}-f''(j)\hat{l}^2
\end{split}\end{align}
where
\begin{equation}
    f(j)=(1-j)\log\left(\frac{1-j}{2}\right)+(1+j)\log\left(\frac{1+j}{2}\right).
\end{equation}
For the mean field analysis, we will only keep the leading $\propto N$ term. We then treat the rest of operators in the Hamiltonian as classical variables, i.e. $\hat{c}\to c\sqrt{N}$ and $(\hat{J}_x,0,\hat{J}_z)\to Nj(\sin\theta,0,\cos\theta)/2$ (we assume beforehand that the spin will have $J_y^{\text{mf}}=0$). At order $\propto N$, the resulting free energy $F$ is given by
\begin{align}\begin{split}
    \frac{F}{N}&=\omega |c|^2-\frac{\omega_0j\cos\theta}{2}+\frac{f(j)}{2\beta}+\lambda j\sin\theta(c+\bar{c})
\end{split}\end{align}
Minmizing with respect to $c,\theta,j$ leads to two type of solutions. The first type has $\theta=c=0$ and
\begin{align}\begin{split}
    j&=\tanh\left(\frac{\beta\omega_0}{2}\right)\\
    \frac{F}{N}&=-\frac{1}{\beta}\log\big[2\cosh(\beta\omega_0/2)\big]
\end{split}\end{align}
In this configuration, the system has no cavity field, the spin is pointing along $+z$ and the spin length is determined by the temperature. The free energy is that of $N$ independent two-level systems. This solution describes the disordered phase. The second type of solution has 
\begin{align}\begin{split}\label{eqn:ThermalBehaviour:MeanFieldSuperradiant}
    \frac{\omega\omega_0}{4\lambda^2\cos\theta}&=\tanh\left(\frac{\beta\omega_0}{2\cos\theta}\right)\\
    j&=\tanh\left(\frac{\beta\omega_0}{2\cos\theta}\right)\\
    c&=-\frac{\lambda j\sin\theta}{\omega}\\[5pt]
    \frac{F}{N}&=\frac{\omega_0j}{4}\frac{(\sin\theta)^2}{\cos\theta}-\frac{1}{\beta}\log\left[2\cosh\left(\frac{\beta\omega_0}{2\cos\theta}\right)\right]
\end{split}\end{align}
The first equation determines the rotation angle as a function of temperature and Hamiltonian parameters, while the last two provide the associated values of spin length and cavity field. There are two possible solutions to these equations, related by $\theta\to-\theta$ and $c\to-c$. These configurations correspond to the ordered superradiant phase and only exist when $\lambda_{\text{eff}}^2=\lambda^2j>\omega\omega_0/4$. This determines the critical temperature $\tanh(\beta_c\omega_0/2)=\omega\omega_0/(4\lambda^2)$ below which the system orders. This approach to the thermal Dicke transition provides a very intuitive picture of the underlying physics: the primary effect of a finite temperature is to establish an equilibrium spin length $j$ via a competition between energy an entropy. Once $j$ is fixed, this univocally determines whether the low energy spectrum of $\hat{H}_{D}$ displays symmetry-breaking or not.

Applying the type I replacement rules for the collective spin operators allows us to obtain the effective Hamiltonian  that describes fluctuations of the system. In the disordered phase the spin is already pointing along $+z$, so we can use the rules directly. As in previous examples terms proportional to $\sqrt{N}$ cancel, leading to
\begin{equation}\label{eqn:ThermalBejaviour:EffectiveH}
    \frac{\hat{K}_{\text{eff}}}{\beta}=\omega_0\Adag\Aop+\omega\hat{c}^\dagger\hat{c}+\lambda_{\text{eff}}(\Adag+\Aop)(\hat{c}+\hat{c}^\dagger)+\frac{f''(j)}{\beta}\hat{l}^2
\end{equation}
Longitudinal fluctuations decouple from the other degrees of freedom, and have size $\delta J\sim \sqrt{N}$. Transverse fluctuations couple to the cavity field, but the effective coupling constant $\lambda_{\text{eff}}=\lambda\sqrt{j}\leq\lambda$ is temperature dependent and becomes larger with decreasing temperature~\cite{PerezFernandez2017}. If $\lambda>\sqrt{\omega\omega_0}/2$ the system develops an instability when $\lambda_{\text{eff}}=\lambda\sqrt{j}=\sqrt{\omega\omega_0}/2$. If $\lambda<\sqrt{\omega\omega_0}/2$, the system cannot reach the instability for any temperature. Equation~(\ref{eqn:ThermalBejaviour:EffectiveH}) can also be used to calculate the excitation spectrum at finite temperature, and compute average values and correlation functions using $e^{-\hat{K}_{\text{eff}}}$ as the approximate quantum state (summing over both mean field solutions in the case of the superradiant phase).

In the superradiant phase we first need to rotate the spin operators and displace the cavity boson before applying the replacement rules. The resulting effective Hamiltonian is instead (with displaced cavity field $\hat{d}=\hat{c}-\sqrt{N}c$)
\begin{align}\begin{split}
    \hat{K}_{\text{eff}}^{\text{sr}}&=\beta\omega\,\hat{d}^\dagger\hat{d}+\frac{\beta\omega_0}{\cos\theta}\Adag\Aop+f''(j)\,\hat{l}^2\\
    &+\beta\lambda(\hat{d}+\hat{d}^\dagger)\left[\sqrt{j}\cos\theta(\Aop+\Adag)+2\sin\theta\,\hat{l}\right],
\end{split}\end{align}
and indicates that the spin length fluctuations now couple to the rest of degrees of freedom. Both effective Hamiltonians for the Dicke model at finite temperature were derived before in Ref.~\cite{Alavirad2019} using diagrammatic methods and a fermionic Majorana representation of spin $1/2$ systems. Our method provides the same results, but our variables of choice ($\Aop,\hat{l}$) possess intrinsic geometric meaning.

We finalize this section by studying the phase transition region. As in all the previous examples, we begin from the disordered phase. As the temperature is decreased and $\lambda_{\text{eff}}$ approaches the critical value, one of the normal modes of the system becomes soft (its excitation energy goes to $0$ in Eq.~(\ref{eqn:ThermalBejaviour:EffectiveH})], while the other one retains a gap $\sim N^0$. As in Sec.~\ref{subsec:Motivation}, further terms in the $1/N$ expansion will introduce a quartic nonlinearity that creates a gap to excitations of the soft mode of size $\sim N^{-1/3}$. Because of the finite $T_c$, excitations of the gapped mode might be present or not depending on the relative sizes of $T_c$ and other scales of the system such as $\omega,\omega_0$. However, the soft mode will always be highly excited, and can be treated classically. Furthermore, the soft mode will turn out to couple nonlinearly to fluctuations in the spin length, which can also be treated classically because $e^{-\beta\hat{H}_D}$ is always diagonal in the $\hat{J}$ basis. 

\begin{figure}
    \centering
    \includegraphics[width=0.98\linewidth]{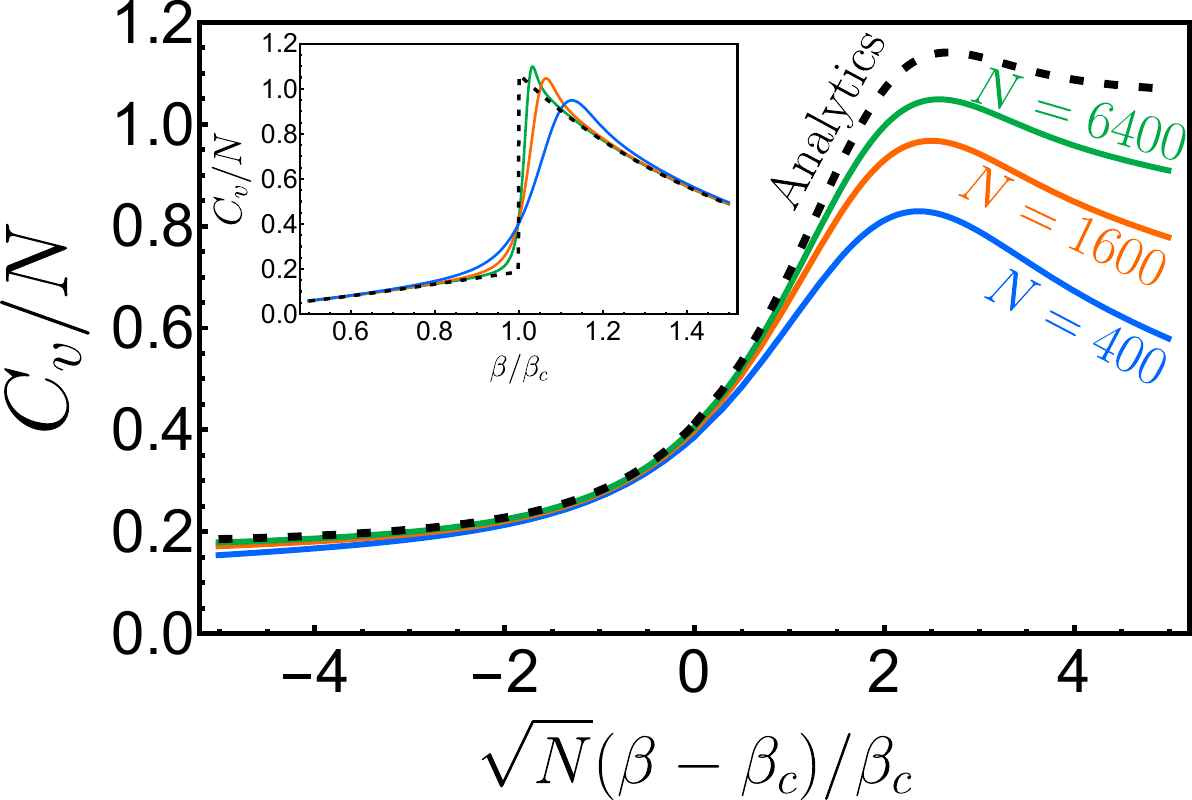}
    \caption{Specific heat $C_v$ as a function of scaled temperature $\sqrt{N}(\beta-\beta_c)/\beta_c$ for different $N=400,1600,6400$ in the LMG limit ($\omega/\omega_0\to\infty$ while keeping $\lambda^2/\omega$ fixed) and fixed $j_c=1/2\to\beta_c\omega_0\approx1.099$. Inset shows the same plot but as a function of $\beta/\beta_c$ illustrating that mean field is a good approximation away from $\beta_c$.}
    \label{fig:thermal}
\end{figure}

Because of these considerations, the effective Hamiltonian in the vicinity of the phase transition will be a combination of a quantum quadratic piece, describing the gapped mode, and a classical nonlinear part, describing the soft and spin length modes. This effective Hamiltonian is given by (see Appendix~\ref{app:Thermal} and omitting constant contributions)
\begin{align}\begin{split}\label{eqn:Thermal:EffectiveHCriticalPoint}
    \hat{K}_{\text{eff}}^{\text{tr}}&=\frac{\beta_c(\omega_0^2+\omega^2)^{1/2}}{2}(\hat{p}_g^2+\hat{g}^2)\\[5pt]
&+ \frac{(\beta_c\omega_0)^{3/2}}{2\sqrt{j}}\left(\frac{\omega^2}{\omega^2+\omega_0^2}\right)p_s^2\\[5pt]
&+f''(j)l^2+\frac{s^4}{4}-\left(\beta_c\omega_0\xi+\sqrt{\frac{\beta_c\omega_0}{j}} s^2\right)l
\end{split}\end{align}
where $[\hat{g},\hat{p}_g]=i$ and $\{\hat{s},\hat{p}_s\}^{\text{pb}}=1$ are the gapped and gapless modes, defined in Eq.~(\ref{eqn:app:Thermal:BosonRelations}) from Appendix~\ref{app:Thermal} and 
\begin{align}\begin{split}
   \xi=\frac{\sqrt{N}(\beta-\beta_c)}{\beta_c}
    % &=-2(Nj)^{2/3}\left[\frac{(\omega_0^2+\omega^2)}{\omega^2}\right]^{1/3}\left(\frac{2\lambda\sqrt{j}}{\sqrt{\omega\omega_0}}-1\right)\\  
\end{split}\end{align}
measures the relative deviation from the critical temperature in units of $1/\sqrt{N}$. Spin observables will include contributions from both soft and gapped modes, which may make a simple finite size scaling analysis more challenging. % \begin{align}\begin{split}\label{eqn:Thermal:EffectiveHCriticalPoint}
%     \hat{H}_{D}^{\text{eff,tr}}&=-\frac{\omega_0Nj_c}{2}-\frac{(\omega+\omega_0)}{2}+\frac{\sqrt{\omega_0^2+\omega^2}}{2}(\hat{p}_g^2+\hat{g}^2)\\
% &+\frac{1}{(Nj_c)^{1/3}}\left[\frac{\omega^4\omega_0^3}{(\omega_0^2+\omega^2)^2}\right]^{1/3}\left(\frac{\hat{p}_s^2}{2}+\frac{\xi\hat{s}^2}{2}+\frac{\hat{s}^4}{4}\right)
% \end{split}\end{align}
% where $[\hat{g},\hat{p}_g]=i$ and $[\hat{s},\hat{p}_s]=i$ are the gapped and gapless modes, defined as
% \begin{align}\begin{split}
%  \hat{g}&=\frac{\omega_0(\frac{\omega}{\omega_0})^{1/4}(\Aop+\Adag)+\omega (\frac{\omega_0}{\omega})^{1/4}(\hat{c}+\hat{c}^\dagger)}{\sqrt{2}(\omega_0^2+\omega^2)^{1/2}}\\[10pt]
%     \hat{s}&=\left[\frac{Nj\sqrt{\omega\omega_0}(\omega_0^2+\omega^2)^2}{\omega^3\omega_0^2}\right]^{1/6}\\
%     &\times\left[\frac{\omega(\frac{\omega}{\omega_0})^{1/4}(\Aop+\Adag)-\omega_0(\frac{\omega_0}{\omega})^{1/4}(\hat{c}+\hat{c}^\dagger)}{\sqrt{2}(\omega_0^2+\omega^2)^{1/2}}\right]\\
% \end{split}\end{align}
% and 
% \begin{align}\begin{split}
%     \xi &=(Nj)^{2/3}\left[\frac{(\omega_0^2+\omega^2)}{\omega^2}\right]^{1/3}\frac{\omega_0(\beta_c-\beta)(1-j^2)}{2j}
%     % &=-2(Nj)^{2/3}\left[\frac{(\omega_0^2+\omega^2)}{\omega^2}\right]^{1/3}\left(\frac{2\lambda\sqrt{j}}{\sqrt{\omega\omega_0}}-1\right)\\  
% \end{split}\end{align}
Because of this, we focus on the specific heat ($C_v$) as the phase transition point is crossed, which will also be dominated by the soft mode. We can obtain an analytical expression  for $C_v$
\begin{equation}
\frac{C_v}{N}=\frac{\beta_c^2\omega_0^2(1-j_c^2)}{4}+b^2g''\left(b\xi\right),    
\end{equation}
where $g(z)=\log\left[\int dx\,\exp(-x^4+zx^2)\right]$ can be expressed in terms of modified Bessel functions, and
\begin{equation}
    b=\frac{(\beta_c\omega_0)^{3/2}(1-j_c^2)}{\sqrt{4j_c-2\beta_c\omega_0(1-j_c^2)}}.
\end{equation}
We point out that this contribution to $C_v$ depends only on $\beta_c\omega_0$ (which also determines $j_c$) and $(\beta-\beta_c)/\beta_c$, and is otherwise independent of $\omega/\omega_0$. In particular, the same formula should hold in the limit $\omega\gg\omega_0$ if we keep $\beta_c\omega_0$ [or equivalently $\lambda^2/(\omega\omega_0)$] fixed, in which case the Dicke model reduces to Eq.~(\ref{eqn:Motivation:Hamiltonian}) with $g=4\lambda^2/(\omega\omega_0)$, and for which larger system sizes can be numerically probed. We show the specific heat as a function of $\xi$ in Fig.~\ref{fig:thermal} for $N$ up to $6400$, calculated by brute-force evaluation of the partition sum. There is a good agreement with the analytical formula for $\xi\lesssim 1$. For $\xi\gtrsim 1$ the numerical results have not yet converged to their $N\to\infty$ limit, presumably because we are zooming in on a violent discontinuity (see inset), but the numerical curves seem to be approaching the analytical result.
\section{Conclusions and outlook}
In this paper we have shown in detail how to construct a Schwinger boson mapping for systems of  $N$ spin $1/2$'s undergoing open system, permutationally symmetric, dynamics. Using this mapping we then introduced a generalization of the Holstein-Primakoff transformation and developed a systematic expansion in powers of $1/N$. We explicitly calculated the leading and next-to-leading order terms in the expansion and illustrated how to use it by means of various examples. These examples included the analysis of driven-dissipative and thermal phase transitions and their finite size scaling properties.

We believe these methods have wide applicability and could be helpful in the analysis of the various permutationally symmetric models that are routinely studied in the literature. This includes the various generalizations of the Dicke model that have been considered over the years, but also models arising in other areas of study, such as reaction-diffusion phenomena~\cite{wampler2025absorbingstatephasetransitions} and non-reciprocal interactions~\cite{Nadolny2025}.

Although we developed the Schwinger boson mapping by rewriting the results from Ref.~\cite{Chase2008}, the rotationally covariant structure that we identified in Eq.~(\ref{eqn:DH:GeneralizedSchwinger}) hints at the possibility of a different, simpler, group theoretic derivation. Such a derivation would also be of use when seeking multilevel generalizations of the mapping in the presence of single particle dissipation (in the strongly symmetric case the mapping is standard~\cite{DAvis2019,Rui2022,Bhuvanesh2024}). 

\section{Acknowledgements}
The author thanks N. Cooper for helpful discussions and O. Scarlatella and M. Wampler for feedback on this manuscript. This work was supported by the Simons Investigator Award (Grant No. 511029) and the Engineering and Physical Sciences Research Council [grant numbers EP/V062654/1 and EP/Y01510X/1]. Numerical simulations were performed using QuTIP~\cite{JOHANSSON20131234} and its Permutationally Invariant Quantum Solver (PIQS)~\cite{Shammah2018}.
\bibliography{library}
\appendix

\onecolumngrid
\section{Expressing local dissipators in terms of Schwinger bosons}\label{app:SchwingerDissipators}
In this section, we show that
\begin{equation}\label{eqn:app:Schwinger:OurFormula}
    \sum_{i=1}^N\hat{s}_{\alpha}^i\hat{\rho}\,\hat{s}^i_{\beta}=E(\hat{J})\,\hat{J}_\alpha\hat{\rho}\hat{J}_\beta+F(\hat{J})\,\hat{K}_\alpha\hat{\rho}\hat{L}_\beta+G(\hat{J})\,\hat{L}_\alpha\hat{\rho}\hat{K}_\beta,
\end{equation}
where
\begin{align}\begin{split}
    E(\hat{J})&=\frac{1+N/2}{2\hat{J}(\hat{J}+1)}\\[5pt]
    F(\hat{J})&=\frac{N/2+\hat{J}+2}{2(\hat{J}+1)(2\hat{J}+3)}\\[5pt]
    G(\hat{J})&=\frac{N/2-\hat{J}+1}{2\hat{J}(2\hat{J}-1)}\\[5pt]
    \mathbf{\hat{K}}&=\frac{1}{2}\begin{pmatrix}
        \hat{b}&\hat{a}
    \end{pmatrix}\,i\sigma_y\bm{\sigma}\begin{pmatrix}
        \hat{b}\\ \hat{a}
    \end{pmatrix}\\[5pt]
    \mathbf{\hat{L}}&=-\frac{1}{2}\begin{pmatrix}
        \hat{b}^\dagger&\hat{a}^\dagger
    \end{pmatrix}\bm{\sigma}\,i\sigma_y\begin{pmatrix}
        \hat{b}^\dagger\\ \hat{a}^\dagger
    \end{pmatrix}
\end{split}\end{align}
is equivalent to the matrix elements calculated in Ref.~\cite{Chase2008}. These were given as [Eq.~(42) in Ref.~\cite{Chase2008}]
\begin{align}\begin{split}\label{eqn:app:Schwinger:ChaseFormula}
    \sum_{n=1}^N\hat{s}_q^{(n)}\overline{\ketbra{J,M}{J,M'}}(\hat{s}_r^{(n)})^\dagger&=\frac{1}{2J}\left(1+\frac{\alpha_N^{J+1}}{d^J_N}\frac{2J+1}{J+1}\right)\times A_q^{J,M}\overline{\ketbra{J,M_q}{J,M'_r}}A_{r}^{J,M'}\\[5pt]
    &+\frac{\alpha_N^J}{2Jd_N^J}\times B^{J,M}_q\overline{\ketbra{J-1,M_q}{J-1,M'_r}}B_{r}^{J,M'}\\[5pt]
    &+\frac{\alpha_N^{J+1}}{2(J+1)d_N^J}\times D^{J,M}_q\overline{\ketbra{J+1,M_q}{J+1,M'_r}}D_{r}^{J,M'},
\end{split}\end{align}
We will first explain the various objects that appear in this formula. First, the $\overline{\ketbra{J,M}{J,M'}}$ are the permutationally symmetric density matrices that are right (left) eigenmatrices of $\hat{J}_z$ with eigenvalue $M$ ($M'$) and left/right eigenmatrices of $\hat{J}$ with equal eigenvalue $J$. The specific normalization chosen in Ref.~\cite{Chase2008} will not be relevant for our discussion. Then $q,r$ range over $={+,-,z}$, with $M_{\pm}=M\pm1, M_z=M$ and there are various numerical coefficients
\begin{align}
    \begin{split}
        \alpha^J_N&=\frac{N!}{(N/2-J)!(N/2+J)!}\\
        d^J_N&=\frac{N!(2J+1))}{(N/2-J)!(N/2+J+1)!}\\[10pt] 
A_+^{J,M}&=\sqrt{(J-M)(J+M+1)}\\ A_-^{J,M}&=\sqrt{(J+M)(J-M+1)}\\A_z^{J,M}&=M\\[10pt]
           B_+^{J,M}&=\sqrt{(J-M)(J-M-1)}\\ B_-^{J,M}&=-\sqrt{(J+M)(J+M-1)}\\B_z^{J,M}&=\sqrt{(J+M)(J-M)}\\[10pt]
D_+^{J,M}&=-\sqrt{(J+M+1)(J+M+2)}\\D_-^{J,M}&=\sqrt{(J-M+1)(J-M+2)}\\D_z^{J,M}&=\sqrt{(J+M+1)(J-M+1)}
    \end{split}
\end{align}
Note that Ref.~\cite{Chase2008} wrote $\hat{\sigma}$ instead of $\hat{s}$ in Eq.~(\ref{eqn:app:Schwinger:ChaseFormula}), but the right hand side of the equation matches the expression with spin $1/2$ operators $\hat{s}$ (this can be checked by setting $q=r=z$ and taking traces of both sides). This does not matter for $\hat{\sigma}_\pm=\hat{s}_\pm=\hat{s}_x\pm i\hat{s}_y=(\hat{\sigma}_x\pm i\hat{\sigma}_y)/2$, but is important for $\hat{s}_z=\hat{\sigma}_z/2$. To proceed, let us begin by calculating the coefficients in front of $A_q^{J,M}\ketbra{J,M}{J,M'}A_r^{J,M'}$, $B_q^{J,M}\ketbra{J-1,M}{J,M'}A_r^{J-1,M'}$, $D_q^{J+1,M}\ketbra{J,M}{J,M'}D_r^{J+1,M'}$ which are
\begin{align}\begin{split}
    \frac{1}{2J}\left(1+\frac{\alpha_N^{J+1}}{d^J_N}\frac{2J+1}{J+1}\right)&=\frac{N/2+1}{2J(J+1)}=E(J)\\
    \frac{\alpha_N^J}{2J d_N^J}&=\frac{N/2+J+1}{(2J+1)2J}=F(J-1)\\
    \frac{\alpha_N^{J+1}}{2(J+1)d^J_N}&=\frac{N/2-J}{2(J+1)(2J+1)}=G(J+1).
\end{split}\end{align}
Noting that 
\begin{align}
    \begin{split}
        E(J)\ketbra{J,M}{J,M'}&=E(\hat{J})\ketbra{J,M}{J,M'}\\
        F(J-1)\ketbra{J-1,M}{J-1,M'}&=F(\hat{J})\ketbra{J-1,M}{J-1,M'}\\
        G(J+1)\ketbra{J+1,M}{J+1,M'}&=G(\hat{J})\ketbra{J+1,M}{J+1,M'},     
    \end{split}
\end{align}
we can rewrite Eq.~(\ref{eqn:app:Schwinger:ChaseFormula}) as
\begin{align}\begin{split}\label{eqn:app:Schwinger:ChaseFormula2}
    \sum_{n=1}^N\hat{s}_q^{(n)}\overline{\ketbra{J,M}{J,M'}}(\hat{s}_r^{(n)})^\dagger&=E(\hat{J})\,A_q^{J,M}\overline{\ketbra{J,M_q}{J,M'_r}}A_{r}^{J,M'}+F(\hat{J})\, B^{J,M}_q\overline{\ketbra{J-1,M_q}{J-1,M'_r}}B_{r}^{J,M'}\\[5pt]
    &+G(\hat{J})\, D^{J,M}_q\overline{\ketbra{J+1,M_q}{J+1,M'_r}}D_{r}^{J,M'},
\end{split}\end{align}
which already indicates which terms in Eq.~(\ref{eqn:app:Schwinger:ChaseFormula2}) should be identified with which in Eq.~(\ref{eqn:app:Schwinger:OurFormula}). To proceed, we express $J,M$ and in terms of the Schwinger boson occupation numbers $n_a=J-M,n_b=J+M$ and analyze Eq.~(\ref{eqn:app:Schwinger:ChaseFormula2}) on a case-by-case basis
\begin{itemize}
    \item $\boxed{\sum_{n=1}^N\hat{s}_z^{(n)}\overline{\ketbra{J,M}{J,M'}}\hat{s}_z^{(n)}}$
    \begin{align}\begin{split}
        \sum_{n=1}^N\hat{s}_z^{(n)}\overline{\ketbra{n_a,n_b}{n_a',n_b'}}(\hat{s}_z^{(n)})^\dagger&=E(\hat{J})\,\left(\frac{n_b-n_a}{2}\right)\overline{\ketbra{n_a,n_b}{n_a',n_b'}}\left(\frac{n_b'-n_a'}{2}\right)\\[5pt]
        &+F(\hat{J})\, \sqrt{n_a^{\vphantom{'}}n_b^{\vphantom{'}}}\hspace{0.2cm}\overline{\ketbra{n_a-1,n_b-1}{n_a'-1,n_b'-1}}\hspace{0.2cm}\sqrt{n_a'n_b'}\\[5pt]
    &+G(\hat{J})\, \sqrt{(n_a^{\vphantom{'}}+1)(n_b^{\vphantom{'}}+1)}\hspace{0.2cm}\overline{\ketbra{n_a+1,n_b+1}{n_a'+1,n_b'+1}}\hspace{0.2cm}\sqrt{(n_a'+1)(n_b'+1)}\\[15pt]
    &=E(\hat{J})\,\left(\frac{\hat{b}^\dagger\hat{b}-\hat{a}^\dagger\hat{a}}{2}\right)\overline{\ketbra{n_a,n_b}{n_a',n_b'}}\left(\frac{\hat{b}^\dagger\hat{b}-\hat{a}^\dagger\hat{a}}{2}\right)\\[5pt]
        &+F(\hat{J})\hspace{0.2cm}\hat{a}\hat{b}\hspace{0.2cm}\overline{\ketbra{n_a,n_b}{n_a',n_b'}}\hspace{0.2cm}\hat{a}^\dagger\hat{b}^\dagger\\[5pt]
    &+G(\hat{J})\hspace{0.2cm} \hat{a}^\dagger\hat{b}^\dagger\hspace{0.2cm}\overline{\ketbra{n_a,n_b}{n_a',n_b'}}\hspace{0.2cm}\hat{a}\hat{b}
    \end{split}\end{align}

    \item $\boxed{\sum_{n=1}^N\hat{\sigma}_+^{(n)}\overline{\ketbra{J,M}{J,M'}}\hat{\sigma}_-^{(n)}}$
    \begin{align}\begin{split}
        \sum_{n=1}^N\hat{\sigma}_+^{(n)}\overline{\ketbra{n_a,n_b}{n_a',n_b'}}(\hat{\sigma}_+^{(n)})^\dagger&=E(\hat{J})\hspace{0.2cm}\sqrt{n_a(n_b+1)}\hspace{0.2cm}\overline{\ketbra{n_a-1,n_b+1}{n_a'-1,n_b'+1}}\hspace{0.2cm}\sqrt{n_a'(n_b'+1)}\\[5pt]
        &+F(\hat{J})\hspace{0.2cm}\sqrt{n_a(n_a-1)}\hspace{0.2cm}\overline{\ketbra{n_a-2,n_b}{n_a'-2,n_b'}}\hspace{0.2cm}\sqrt{n_a'(n_a'-1)}\\[5pt]
    &+G(\hat{J})\hspace{0.2cm}\sqrt{(n_b+1)(n_b+2)}\hspace{0.2cm}\overline{\ketbra{n_a,n_b+2}{n_a',n_b'+2}}\hspace{0.2cm}\sqrt{(n_b'+1)(n_b'+2)}\\[15pt]
    &=E(\hat{J})\hspace{0.2cm}\hat{a}\hat{b}^\dagger\hspace{0.2cm}\overline{\ketbra{n_a,n_b}{n_a',n_b'}}\hspace{0.2cm}\hat{a}^\dagger\hat{b}\\[5pt]
        &+F(\hat{J})\hspace{0.2cm}\hat{a}^2\hspace{0.2cm}\overline{\ketbra{n_a,n_b}{n_a',n_b'}}\hspace{0.2cm}(\hat{a}^\dagger)^2\\[5pt]
    &+G(\hat{J})\hspace{0.2cm} (\hat{b}^\dagger)^2\hspace{0.2cm}\overline{\ketbra{n_a,n_b}{n_a',n_b'}}\hspace{0.2cm}\hat{b}^2
    \end{split}\end{align}

    \item $\boxed{\sum_{n=1}^N\hat{\sigma}_-^{(n)}\overline{\ketbra{J,M}{J,M'}}\hat{\sigma}_+^{(n)}}$
    \begin{align}\begin{split}
        \sum_{n=1}^N\hat{\sigma}_-^{(n)}\overline{\ketbra{n_a,n_b}{n_a',n_b'}}(\hat{\sigma}_-^{(n)})^\dagger&=E(\hat{J})\hspace{0.2cm}\sqrt{n_b(n_a+1)}\hspace{0.2cm}\overline{\ketbra{n_a+1,n_b-1}{n_a'+1,n_b'-1}}\hspace{0.2cm}\sqrt{n_b'(n_a'+1)}\\[5pt]
        &+F(\hat{J})\hspace{0.2cm}\sqrt{n_b(n_b-1)}\hspace{0.2cm}\overline{\ketbra{n_a,n_b-2}{n_a',n_b'-2}}\hspace{0.2cm}\sqrt{n_b'(n_b'-1)}\\[5pt]
    &+G(\hat{J})\hspace{0.2cm}\sqrt{(n_a+1)(n_a+2)}\hspace{0.2cm}\overline{\ketbra{n_a+2,n_b}{n_a'+2,n_b'}}\hspace{0.2cm}\sqrt{(n_a'+1)(n_a'+2)}\\[15pt]
    &=E(\hat{J})\hspace{0.2cm}\hat{a}^\dagger\hat{b}\hspace{0.2cm}\overline{\ketbra{n_a,n_b}{n_a',n_b'}}\hspace{0.2cm}\hat{a}\hat{b}^\dagger\\[5pt]
        &+F(\hat{J})\hspace{0.2cm}\hat{b}^2\hspace{0.2cm}\overline{\ketbra{n_a,n_b}{n_a',n_b'}}\hspace{0.2cm}(\hat{b}^\dagger)^2\\[5pt]
    &+G(\hat{J})\hspace{0.2cm} (\hat{a}^\dagger)^2\hspace{0.2cm}\overline{\ketbra{n_a,n_b}{n_a',n_b'}}\hspace{0.2cm}\hat{a}^2
    \end{split}\end{align}

    \item $\boxed{\sum_{n=1}^N\hat{\sigma}_-^{(n)}\overline{\ketbra{J,M}{J,M'}}\hat{\sigma}_-^{(n)}}$
    \begin{align}\begin{split}
        \sum_{n=1}^N\hat{\sigma}_-^{(n)}\overline{\ketbra{n_a,n_b}{n_a',n_b'}}(\hat{\sigma}_+^{(n)})^\dagger&=E(\hat{J})\hspace{0.2cm}\sqrt{n_b(n_a+1)}\hspace{0.2cm}\overline{\ketbra{n_a+1,n_b-1}{n_a'-1,n_b'+1}}\hspace{0.2cm}\sqrt{n_a'(n_b'+1)}\\[5pt]
        &-F(\hat{J})\hspace{0.2cm}\sqrt{n_b(n_b-1)}\hspace{0.2cm}\overline{\ketbra{n_a,n_b-2}{n_a'-2,n_b'}}\hspace{0.2cm}\sqrt{n_a'(n_a'-1)}\\[5pt]
    &-G(\hat{J})\hspace{0.2cm}\sqrt{(n_a+1)(n_a+2)}\hspace{0.2cm}\overline{\ketbra{n_a+2,n_b}{n_a',n_b'+2}}\hspace{0.2cm}\sqrt{(n_b'+1)(n_b'+2)}\\[15pt]
    &=E(\hat{J})\hspace{0.2cm}\hat{a}^\dagger\hat{b}\hspace{0.2cm}\overline{\ketbra{n_a,n_b}{n_a',n_b'}}\hspace{0.2cm}\hat{a}^\dagger\hat{b}\\[5pt]
        &-F(\hat{J})\hspace{0.2cm}\hat{b}^2\hspace{0.2cm}\overline{\ketbra{n_a,n_b}{n_a',n_b'}}\hspace{0.2cm}(\hat{a}^\dagger)^2\\[5pt]
    &-G(\hat{J})\hspace{0.2cm} (\hat{a}^\dagger)^2\hspace{0.2cm}\overline{\ketbra{n_a,n_b}{n_a',n_b'}}\hspace{0.2cm}\hat{b}^2
    \end{split}\end{align}

    \item $\boxed{\sum_{n=1}^N\hat{\sigma}_-^{(n)}\overline{\ketbra{J,M}{J,M'}}\hat{s}_z^{(n)}}$
    \begin{align}\begin{split}
        \sum_{n=1}^N\hat{\sigma}_-^{(n)}\overline{\ketbra{n_a,n_b}{n_a',n_b'}}(\hat{s}_z^{(n)})^\dagger&=E(\hat{J})\hspace{0.2cm}\sqrt{n_b(n_a+1)}\hspace{0.2cm}\overline{\ketbra{n_a+1,n_b-1}{n_a',n_b'}}\hspace{0.2cm}\left(\frac{n_b'-n_a'}{2}\right)\\[5pt]
        &-F(\hat{J})\hspace{0.2cm}\sqrt{n_b(n_b-1)}\hspace{0.5cm}\overline{\ketbra{n_a,n_b-2}{n_a'-1,n_b'-1}}\hspace{0.5cm}\sqrt{n_a'n_b'}\\[5pt]
    &+G(\hat{J})\hspace{0.2cm}\sqrt{(n_a+1)(n_a+2)}\hspace{0.5cm}\overline{\ketbra{n_a+2,n_b}{n_a'+1,n_b'+1}}\hspace{0.5cm}\sqrt{(n_a'+1)(n_b'+1)}\\[15pt]
    &=E(\hat{J})\hspace{0.2cm}\hat{a}^\dagger\hat{b}\hspace{0.2cm}\overline{\ketbra{n_a,n_b}{n_a',n_b'}}\hspace{0.2cm}(\hat{b}^\dagger\hat{b}-\hat{a}^\dagger\hat{a})/2\\[5pt]
        &-F(\hat{J})\hspace{0.2cm}\hat{b}^2\hspace{0.2cm}\overline{\ketbra{n_a,n_b}{n_a',n_b'}}\hspace{0.2cm}\hat{a}^\dagger\hat{b}^\dagger\\[5pt]
    &+G(\hat{J})\hspace{0.2cm} (\hat{a}^\dagger)^2\hspace{0.2cm}\overline{\ketbra{n_a,n_b}{n_a',n_b'}}\hspace{0.2cm}\hat{a}\hat{b}
    \end{split}\end{align}

     \item $\boxed{\sum_{n=1}^N\hat{\sigma}_+^{(n)}\overline{\ketbra{J,M}{J,M'}}\hat{s}_z^{(n)}}$
    \begin{align}\begin{split}
        \sum_{n=1}^N\hat{\sigma}_+^{(n)}\overline{\ketbra{n_a,n_b}{n_a',n_b'}}(\hat{s}_z^{(n)})^\dagger&=E(\hat{J})\hspace{0.2cm}\sqrt{n_a(n_b+1)}\hspace{0.2cm}\overline{\ketbra{n_a-1,n_b+1}{n_a',n_b'}}\hspace{0.2cm}\left(\frac{n_b'-n_a'}{2}\right)\\[5pt]
        &+F(\hat{J})\hspace{0.2cm}\sqrt{n_a(n_a-1)}\hspace{0.5cm}\overline{\ketbra{n_a-2,n_b}{n_a'-1,n_b'-1}}\hspace{0.5cm}\sqrt{n_a'n_b'}\\[5pt]
    &-G(\hat{J})\hspace{0.2cm}\sqrt{(n_b+1)(n_b+2)}\hspace{0.5cm}\overline{\ketbra{n_a,n_b+2}{n_a'+1,n_b'+1}}\hspace{0.5cm}\sqrt{(n_a'+1)(n_b'+1)}\\[15pt]
    &=E(\hat{J})\hspace{0.2cm}\hat{a}^\dagger\hat{b}\hspace{0.2cm}\overline{\ketbra{n_a,n_b}{n_a',n_b'}}\hspace{0.2cm}(\hat{b}^\dagger\hat{b}-\hat{a}^\dagger\hat{a})/2\\[5pt]
        &+F(\hat{J})\hspace{0.2cm}\hat{a}^2\hspace{0.2cm}\overline{\ketbra{n_a,n_b}{n_a',n_b'}}\hspace{0.2cm}\hat{a}^\dagger\hat{b}^\dagger\\[5pt]
    &-G(\hat{J})\hspace{0.2cm} (\hat{b}^\dagger)^2\hspace{0.2cm}\overline{\ketbra{n_a,n_b}{n_a',n_b'}}\hspace{0.2cm}\hat{a}\hat{b}
    \end{split}\end{align}
\end{itemize}

These expressions agree with Eq.~(\ref{eqn:app:Schwinger:OurFormula}), taking into account that $\hat{K}_{\pm}=\hat{K}_{x}\pm i\hat{K}_{y}$, $\hat{L}_{\pm}=\hat{L}_{x}\pm i\hat{L}_{y}$ and $\hat{s}_z=\hat{\sigma}_z/2$.

\section{Derivation of replacement rules}\label{app:ReplacementRuleDerivation}
In this appendix we derive the replacement rules provided in Table~\ref{tab:typLind} and Table~\ref{tab:typLindII}. We begin from the results in Sec.~\ref{app:SchwingerDissipators} and get rid of $\hat{b}$ in favor of $\hat{J}$. To do this, we decompose the boson $\hat{b}$ using its number-phase representation $\hat{b}=e^{i\hat{\phi}/2}(\hat{b}^\dagger\hat{b})^{1/2}$, absorb $e^{i\hat{\phi}/2}$ into $\hat{A}^\dagger=\hat{a}^\dagger e^{i\hat{\phi}/2}$, replace $\hat{b}^\dagger\hat{b}=2\hat{J}-\Adag\Aop$. Then we obtain, generically

\begin{itemize}
    \item $\boxed{\sum_{i=1}^N\hat{s}_z^i\hat{\rho}\,\hat{s}_z^i}$
    \begin{align}\begin{split}
        \sum_{i=1}^N\hat{s}_z\hat{\rho}\,\hat{s}_z&=E(\hat{J})\left(\hat{J}-\Adag\Aop\right)\hat{\rho}\left(\hat{J}-\Adag\Aop\right)\\
        &+F(\hat{J})\Aop \left(2\hat{J}+2-\Adag\Aop\right)^{1/2}\ephi\hat{\rho}\enphi\left(2\hat{J}+2-\Adag\Aop\right)^{1/2}\Adag\\
        &+G(\hat{J})\left(2\hat{J}-\Adag\Aop\right)^{1/2}\enphi\Adag\hat{\rho}\Aop\ephi\left(2\hat{J}-\Adag\Aop\right)^{1/2}
    \end{split}\end{align}

    \item $\boxed{\sum_{i=1}^N\hat{\sigma}_+^i\hat{\rho}\,\hat{\sigma}_-^i}$
    \begin{align}\begin{split}
       \sum_{i=1}^N\hat{\sigma}_+^i\hat{\rho}\,\hat{\sigma}_-^i&=E(\hat{J})\left(2\hat{J}-\Adag\Aop\right)^{1/2}\Aop\hat{\rho}\Adag\left(2\hat{J}-\Adag\Aop\right)^{1/2}\\
        &+F(\hat{J})(\Aop)^2 \ephi\hat{\rho}\enphi(\Adag)^2\\
        &+G(\hat{J})\left(2\hat{J}-\Adag\Aop\right)^{1/2}\left(2\hat{J}-1-\Adag\Aop\right)^{1/2}\enphi\hat{\rho}\ephi\left(2\hat{J}-\Adag\Aop\right)^{1/2}\left(2\hat{J}-1-\Adag\Aop\right)^{1/2}
    \end{split}\end{align}

        \item $\boxed{\sum_{i=1}^N\hat{\sigma}_-^i\hat{\rho}\,\hat{\sigma}_+^i}$
    \begin{align}\begin{split}
       \sum_{i=1}^N\hat{\sigma}_-^i\hat{\rho}\,\hat{\sigma}_+^i&=E(\hat{J})\Adag\left(2\hat{J}-\Adag\Aop\right)^{1/2}\hat{\rho}\left(2\hat{J}-\Adag\Aop\right)^{1/2}\Aop\\
       &+F(\hat{J})\left(2\hat{J}+2-\Adag\Aop\right)^{1/2}\left(2\hat{J}+1-\Adag\Aop\right)^{1/2}\ephi\hat{\rho}\enphi\left(2\hat{J}+2-\Adag\Aop\right)^{1/2}\left(2\hat{J}+1-\Adag\Aop\right)^{1/2}\\
       &+G(\hat{J})(\Adag)^2 \enphi\hat{\rho}\ephi(\Aop)^2
     \end{split}\end{align}

     \item $\boxed{\sum_{i=1}^N\hat{\sigma}_-^i\hat{\rho}\,\hat{\sigma}_-^i}$
    \begin{align}\begin{split}
       \sum_{i=1}^N\hat{\sigma}_-^i\hat{\rho}\,\hat{\sigma}_-^i&=E(\hat{J})\Adag\left(2\hat{J}-\Adag\Aop\right)^{1/2}\hat{\rho}\Adag\left(2\hat{J}-\Adag\Aop\right)^{1/2}\\
       &-F(\hat{J})\left(2\hat{J}+2-\Adag\Aop\right)^{1/2}\left(2\hat{J}+1-\Adag\Aop\right)^{1/2}\ephi\hat{\rho}\enphi(\Adag)^2\\
       &-G(\hat{J})(\Adag)^2\enphi\hrho \ephi\left(2\hat{J}-\Adag\Aop\right)^{1/2}\left(2\hat{J}-1-\Adag\Aop\right)^{1/2}
     \end{split}\end{align}

     \item $\boxed{\sum_{i=1}^N\hat{\sigma}_-^i\hat{\rho}\,\hat{s}_z^i}$
    \begin{align}\begin{split}
       \sum_{i=1}^N\hat{\sigma}_-^i\hat{\rho}\,\hat{s}_z^i&=E(\hat{J})\Adag\left(2\hat{J}-\Adag\Aop\right)^{1/2}\hat{\rho}\left(\hat{J}-\Adag\Aop\right)\\
       &-F(\hat{J})\left(2\hat{J}+2-\Adag\Aop\right)^{1/2}\left(2\hat{J}+1-\Adag\Aop\right)^{1/2}\ephi\hat{\rho}\enphi\left(2\hat{J}+2-\Adag\Aop\right)^{1/2}\Adag\\
       &+G(\hat{J})(\Adag)^2\enphi\hrho \Aop\ephi\left(2\hat{J}-\Adag\Aop\right)^{1/2}
     \end{split}\end{align}

      \item $\boxed{\sum_{i=1}^N\hat{\sigma}_+^i\hat{\rho}\,\hat{s}_z^i}$
    \begin{align}\begin{split}
       \sum_{i=1}^N\hat{\sigma}_+^i\hat{\rho}\,\hat{s}_z^i&=E(\hat{J})\left(2\hat{J}-\Adag\Aop\right)^{1/2}\Aop\hat{\rho}\left(\hat{J}-\Adag\Aop\right)\\
       &+F(\hat{J})(\Aop)^2\ephi\hat{\rho}\enphi\left(2\hat{J}+2-\Adag\Aop\right)^{1/2}\Adag\\
       &-G(\hat{J})\left(2\hat{J}-\Adag\Aop\right)^{1/2}\left(2\hat{J}-1-\Adag\Aop\right)^{1/2}\enphi\hrho \Aop\ephi\left(2\hat{J}-\Adag\Aop\right)^{1/2}
     \end{split}\end{align}
\end{itemize}
These expressions are exact, as is the Holstein-Primakoff mapping for collective states, but full rotational invariance is no longer manifest. As described in the main text, this representation is particularly convenient when the state is polarized along $+z$, but the nature of the expansion will depend on the mean field value of the Bloch vector length. To get Table~\ref{tab:typLind}, valid when $j<1$ (where $j$ is the mean field length in units of $N/2$), we introduce scaled variables $\hat{l},\hat{q}\sim 1$ as follows
\begin{align}
    \begin{split}
        \hat{J}&=\frac{Nj}{2}+\sqrt{N}\,\hat{l}\\
        \hat{\phi}&=\frac{\hat{q}}{\sqrt{N}}
    \end{split}
\end{align}
and expand the expressions to $O(N^0)$, with the neglected terms being of size $N^{-1/2}$. This yields
\begin{itemize}
    \item $\boxed{\sum_{i=1}^N\hat{s}_z^i\hat{\rho}\,\hat{s}_z^i}$
    \begin{align}\begin{split}
        \sum_{i=1}^N\hat{s}_z\hat{\rho}\,\hat{s}_z&=\frac{N}{4}\hrho+\frac{\hrho}{2}-\frac{1}{2j}\Adag\Aop\hrho-\frac{1}{2j}\hrho\left(\Adag\Aop+1\right)+\left(\frac{1+j}{2j}\right)\Aop\hrho\Adag+\left(\frac{1-j}{2}\right)\Adag\hrho\Aop\\[10pt]
        &=\boxed{\frac{N}{4}\hrho+\left(\frac{1+j}{2j}\right)\left(\Aop\rho\Adag-\frac{\{\Aop\Adag,\hrho\}}{2}\right)+\left(\frac{1-j}{2j}\right)\left(\Adag\hrho\Aop+\frac{\{\Aop\Adag,\hrho\}}{2}\right)}
    \end{split}\end{align}
   
    \item $\boxed{\sum_{i=1}^N\hat{\sigma}_+^i\hat{\rho}\,\hat{\sigma}_-^i}$
    \begin{align}\begin{split}
       \sum_{i=1}^N\hat{\sigma}_+^i\hat{\rho}\,\hat{\sigma}_-^i&=\frac{1}{j}\Aop\hrho\Adag+0-\frac{1-j}{2j}\{\Adag\Aop,\hrho\}+\frac{N(1-j)}{2}\hrho-\frac{i\sqrt{N}}{2}(1-j)[\hat{q},\hrho]-\frac{1-j}{4}[\hat{q},[\hat{q},\hrho]]-\sqrt{N}\hat{l}\hrho+i[\hat{q},\hat{l}\hrho]\\[10pt]
        &=\boxed{\frac{N(1-j)\hrho}{2}-\sqrt{N}\left(\frac{i(1-j)}{2}[\hat{q},\hrho]+\hat{l}\hrho\right)+\frac{1}{j}\Aop\hrho\Adag-\frac{(1-j)}{2j}\{\Adag\Aop,\hrho\}+\frac{(1-j)}{4}[\hat{q},[\hrho,\hat{q}]]+i[\hat{q},\hat{l}\hrho]}
    \end{split}\end{align}

        \item $\boxed{\sum_{i=1}^N\hat{\sigma}_-^i\hat{\rho}\,\hat{\sigma}_+^i}$
    \begin{align}\begin{split}
       \sum_{i=1}^N\hat{\sigma}_-^i\hat{\rho}\,\hat{\sigma}_+^i&=\frac{1}{j}\Adag\hrho\Aop+0-\frac{1+j}{2j}\{\Aop\Adag,\hrho\}+\frac{N(1+j)}{2}\hrho+\frac{i\sqrt{N}}{2}(1+j)[\hat{q},\hrho]-\frac{1+j}{4}[\hat{q},[\hat{q},\hrho]]+\sqrt{N}\hat{l}\hrho+i[\hat{q,}\hat{l}\hrho]+\frac{\hrho}{2}\\[10pt]
        &=\boxed{\frac{N(1+j)\hrho}{2}+\sqrt{N}\left(\frac{i(1+j)}{2}[\hat{q},\hrho]+\hat{l}\hrho\right)+\frac{1}{j}\Adag\hrho\Aop-\frac{(1+j)}{2j}\{\Aop\Adag,\hrho\}+\frac{(1+j)}{4}[\hat{q},[\hrho,\hat{q}]]+i[\hat{q},\hat{l}\hrho]+\frac{\hrho}{2}}
     \end{split}\end{align}

     \item $\boxed{\sum_{i=1}^N\hat{\sigma}_-^i\hat{\rho}\,\hat{\sigma}_-^i}$
    \begin{align}\begin{split}
       \sum_{i=1}^N\hat{\sigma}_-^i\hat{\rho}\,\hat{\sigma}_-^i&=\frac{1}{j}\Adag\hrho\Adag-\frac{(1+j)}{2j}\hrho(\Adag)^2-\frac{(1-j)}{2j}(\Adag)^2\hrho\\[10pt]
        &=\boxed{\frac{1}{j}\left((\Adag)^2\hrho(\Adag)^2-\frac{\{(\Adag)^2,\hrho\}}{2}\right)+\frac{1}{2}\big[(\Adag)^2,\hrho\big]}
     \end{split}\end{align}

     \item $\boxed{\sum_{i=1}^N\hat{\sigma}_-^i\hat{\rho}\,\hat{s}_z^i}$
    \begin{align}\begin{split}
       \sum_{i=1}^N\hat{\sigma}_-^i\hat{\rho}\,\hat{s}_z^i&=\frac{1}{2}\sqrt{\frac{N}{j}}\Adag\hrho-\frac{\hat{l}}{2j^{3/2}}\Adag\hrho-\frac{\sqrt{N}(1+j)}{2\sqrt{j}}\hrho\Adag+\frac{(1-j)}{2j^{3/2}}\hat{l}\hrho\Adag-\frac{i(1+j)}{2\sqrt{j}}[\hat{q},\hrho]\Adag\\[10pt]
        &=\boxed{\frac{\sqrt{N}}{2\sqrt{j}}\left(\Adag\hrho-(1+j)\hrho\Adag\right)+\frac{\hat{l}}{2j^{3/2}}\left((1-j)\hrho\Adag-\Adag\hrho\right)-\frac{i(1+j)}{2\sqrt{j}}[\hat{q},\hrho]\Adag}
     \end{split}\end{align}

      \item $\boxed{\sum_{i=1}^N\hat{\sigma}_+^i\hat{\rho}\,\hat{s}_z^i}$
    \begin{align}\begin{split}
       \sum_{i=1}^N\hat{\sigma}_+^i\hat{\rho}\,\hat{s}_z^i&=\frac{1}{2}\sqrt{\frac{N}{j}}\Aop\hrho-\frac{\hat{l}}{2j^{3/2}}\Aop\hrho-\frac{\sqrt{N}(1-j)}{2\sqrt{j}}\hrho\Aop+\frac{(1+j)}{2j^{3/2}}\hat{l}\hrho\Aop+\frac{i(1-j)}{2\sqrt{j}}[\hat{q},\hrho]\Adag\\[10pt]
        &=\boxed{\frac{\sqrt{N}}{2\sqrt{j}}\left(\Aop\hrho-(1-j)\hrho\Aop\right)+\frac{\hat{l}}{2j^{3/2}}\left((1+j)\hrho\Aop-\Aop\hrho\right)+\frac{i(1-j)}{2\sqrt{j}}[\hat{q},\hrho]\Adag}
     \end{split}\end{align}
\end{itemize}

To get Table~\ref{tab:typLindII}, valid when $j=1$, we instead expand the expressions assuming that $\Aop,\Adag,\dJ,\hat{\phi}\sim 1$.

\section{Superradiant laser below upper threshold}\label{app:SuperradiantLaserAboveThreshold}
In this appendix we analyze the superradiant laser model of Eq.~(\ref{eqn:Examples:SuperradiantLasing:BosonMasterEquation}) in the coherent phase. As mentioned in the main text, in this case the system will develop a nonzero $J_-^{\text{mf}}$ in the steady state, which can be chosen to have any arbitrary phase. We thus choose $J_-^{\text{mf}}$ to be real and positive (thus the Bloch vector points along the $x$ direction).  We thus perform a rotation of the spin operators
\begin{equation}
    \hat{\sigma}_+^i=(\hat{s}_z^i)'\sin\theta+(\hat{\sigma}_+^i)'\frac{(1+\cos\theta)}{2}+(\hat{\sigma}_-^i)'\frac{(\cos\theta-1)}{2},
\end{equation}
where $\theta$ is the rotation angle about the $+y$ axis [and hence $\hat{s}_y^i=(\hat{s}_y^i)'$]. In principle $\theta$ is determined from the solution to the mean field equations, but we show in this appendix that it can also be determined by requiring that the terms proportional to $\sqrt{N}$ coming from Table~\ref{tab:typLind} vanish. If we perform the replacement rules in the rotated coordinate system, we arrive at the following bosonic master equation
\begin{align}\begin{split}
    \partial_t\hrho&=\frac{w\sqrt{N}(2-j\cos\theta)\sin\theta}{4\sqrt{j}}\left[\left(\hat{A}-\Adag\right),\hrho\right]+\frac{NC\gamma j\sqrt{Nj}\sin\theta}{4}[\left(\Adag-\Aop\right),\hrho]\\
    &+\frac{iw\sqrt{N}}{8}(1+j)(\cos\theta-1)^2[\hat{q},\hrho]-\frac{iw\sqrt{N}}{8}(1-j)(\cos\theta+1)^2[\hat{q},\hrho]\\[10pt]
    &+\frac{w}{4j}\mathcal{D}\left[(1+\cos\theta)\Aop+(\cos\theta-1)\Adag\right]\hrho+\frac{NC\gamma\, j}{4}\mathcal{D}\left[(1+\cos\theta)\Adag+(\cos\theta-1)\Aop\right]\hrho+\frac{3NC\gamma j^{1/2}\sin\theta}{4}\left[\Adag-\Aop,\hrho\right]\hat{l}\\
    &+\frac{w(j+1)(\sin\theta)^2}{2j}\mathcal{D}[\Aop]\hrho+\frac{w(1-j)(\sin\theta)^2}{2j}\mathcal{D}[\Adag]\hrho+\frac{w(\sin\theta)^2}{4}\left[(\Aop)^2-(\Adag)^2,\hrho\right]\\[10pt]
    &+\frac{w(1-j)(1+\cos\theta)^2+w(1+j)(1-\cos\theta)^2}{8}\mathcal{D}[\hat{q}]\hrho+\frac{iw(1+(\cos\theta)^2)}{4}\big[\hat{q},\{\hat{l},\hrho\}\big]\\
    &-\frac{w\sin\theta(2+j\cos\theta)}{4j^{3/2}}\left[\Aop-\Adag,\hrho\right]\hat{l}+\frac{iw\sin\theta(1-j\cos\theta)}{4\sqrt{j}}\left[\hat{q},\left\{(\Aop+\Adag),\hrho\right\}\right]+\frac{iw\sin\theta(\cos\theta-j)}{4\sqrt{j}}\left[[\hat{q},\hrho],\left(\Aop-\Adag\right)\right]
\end{split}\end{align}
Cancellation of the terms proportional to $\sqrt{N}$ leads to
\begin{equation}
    j(\cos\theta)^2+j=2\cos\theta,\hspace{2cm} NC\gamma j^2=w(2-j\cos\theta)\hspace{2cm} 
\end{equation}
which can be solved to give $j\cos\theta=w/NC\gamma$ (i.e. the $z$ component of the Bloch vector) and $(j\sin\theta)^2=2w(NC\gamma-w)/(NC\gamma)^2$ (i.e. the transverse component of the Bloch vector and is proportional to the emitted light intensity). These are the same results that would be obtained by solving the mean field equations of motion. Massaging this result leads to

\begin{align}\begin{split}
    \partial_t\hrho&=\frac{iw(\sin\theta)^2}{8}\Big(\left[\hat{p},\{\hat{x},\hrho\}\right]+\left[\{\hat{p},\hrho\},\hat{x}\right]\Big)-\frac{iw(\sin\theta)^2}{8}\left[\hat{x}\hat{p}+\hat{p}\hat{x},\hrho\right]+\frac{iw(1+(\cos\theta)^2)}{4}\big[\hat{q},\{\hat{l},\hrho\}\big]\\[5pt]
    &-\frac{w\sin\theta (2+(\cos\theta)^2)}{\sqrt{2j}\cos\theta}\left[\hat{p},\hrho\right]\hat{l}+\frac{iw(\sin\theta)^3\sqrt{j}}{4\sqrt{2}\cos\theta}\left[\hat{q},\{\hrho,\hat{x}\}\right]\\[15pt]
    &+\frac{NC\gamma\cos\theta}{2}(1+j\cos\theta)\mathcal{D}[\hat{x}]\hrho+\left[\left(\frac{w}{2j}+\frac{NC\gamma j}{2}\right)+\frac{w(\sin\theta)^2}{2j}\right]\mathcal{D}[\hat{p}]\hrho\\[5pt]
    &+\frac{wj(\sin\theta)^4}{8\cos\theta}\mathcal{D}[\hat{q}]\hrho+\frac{w(\sin\theta)^3\sqrt{j}}{4\sqrt{2}}\left[\hat{q},\left[\hrho,\hat{p}\right]\right]
\end{split}\end{align}
The first two lines describe relaxation (in accordance with linear response calculated by, e.g., Heisenberg-Langevin equations) and the last two lines describe diffusion. In particular
\begin{equation}
    \frac{d\braket{\hat{p}^2}}{dt}=\frac{NC\gamma\cos\theta}{2}(1+j\cos\theta),
\end{equation}
which is related to the laser phase by $\hat{\phi}=\hat{J}_y/(Nj\sin\theta/2)\approx \hat{p}/(\sin\theta\sqrt{Nj/2})$. Thus, the phase diffuses according to
\begin{equation}
    \frac{1}{2}\frac{d\braket{\hat{\phi}^2}}{dt}=\frac{C\gamma\cos\theta}{2j(\sin\theta)^2}(1+j\cos\theta)=\frac{C\gamma}{4}\left(\frac{NC\gamma+w}{NC\gamma-w}\right),
\end{equation}
which agrees with known results~\cite{tieri2017theorycrossoverlasingsteady} in the appropriate limit ($NC\gamma,w\gg \gamma$) and determines the laser linewidth.

\section{Transverse field Ising model with finite dissipation}\label{app:FiniteGamma}
In this appendix we analyze the transverse field Ising model with finite dissipation. For completeness, we copy here the master equation defining the evolution
\begin{equation}
    \partial_t\hrho=-i\left[-\delta\hat{J}_z-\frac{g}{N}\hat{J}_x^2,\hrho\right]+\gamma\sum_{i=1}^N\left(\hat{\sigma}_i^+\hrho\hat{\sigma}_-^i-\frac{\{\hat{\sigma}_i^-\hat{\sigma}_i^+,\hrho\}}{2}\right)
\end{equation}
In the quadratic approximation, we have Eq.~(\ref{eqn:DDPhaseTransitions:TFIMMasterEquation1})
\begin{align}\label{eqn:app:TFIMMasterEquation1}
    \begin{split}
        \partial_t\hrho&=-\frac{i}{2}\left[\Delta\hat{p}^2+(\Delta-g)\hat{x}^2,\hrho\right]+\gamma\mathcal{D}[\Aop]\hrho+\Delta\left[\dJ,\hrho\right]+\gamma\left(\dJ\hrho-\enphi\dJ\ephi\,\right),
    \end{split}
\end{align}
There is an instability at $\Delta^*$ defined by $\gamma=2\sqrt{\Delta^*(g-\Delta^*)}$. Since the longitudinal boson just equilibrates to $\ket{\delta J=0}$ we project out this degree of freedom and work with the reduced density matrix for the transverse boson $\hrhoT$. If we introduce the conjugate pair $[\hat{x}_s,\hat{p}_f]=i$ according to
\begin{equation}
    \hat{x}_s=\frac{u^{-1}\hat{x}+u\,\hat{p}}{\sqrt{2}},\hspace{1cm}\hat{p}_f=\frac{u\,\hat{p}-u^{-1}\hat{x}}{\sqrt{2}},
\end{equation}
where $u=[\Delta^*/(g-\Delta^*)]^{1/4}$, the equation simplifies to
\begin{equation}
    \partial_t\hrhoT=-\frac{i\gamma}{2}\left[\hat{x}_s,\{\hat{p}_f,\hrhoT\}\right]+\frac{g}{4}\Big(\big[\hat{x}_s,[\hrhoT,\hat{x}_s]\big]+\big[\hat{p}_f,[\hrhoT,\hat{p}_f]\big]\Big)+\left(\frac{g-2\Delta^*}{2}\right)\left[\hat{x}_s,[\hrhoT,\hat{p}_f]\right].
\end{equation}
The first term introduces relaxation for $\hat{p}_f$ with rate $\gamma$ but does not affect $\hat{x}_s$. The next term introduces noise and diffusion, which only manifests in $\braket{\hat{x}_s^2}$ and $\braket{\hat{p}_f^2}$. The last term introduces mixed noise, whch appears in $\braket{\{\hat{x}_s,\hat{p}_f\}}$. Because of the relaxation, the variance of $\hat{p}_f$ never grows too much. However, the noise in $\hat{x}_s$ keeps growing and is stabilized by nonlinearities. We thus introduce $\hat{y}_s=N^{f_x}\hat{x}_s$, which will behave classically, and hence commutators become Poisson brackets according to $[\,,\,]\approx iN^{-f_x}\{\,,\,\}^{\text{pb}}$. The density matrix $\hat{\rho}_T$ becomes a classical probability distribution $\rho_c$ that satisfies the classical master equation
\begin{equation}
    \partial_t\rho_c=\gamma\partial_{p_f}(p_f\,\rho_c)+\frac{g}{4}\partial_{p_f}^2\rho_c+\frac{g}{4N^{2f_x}}\partial_y^2\rho_c-\left(\frac{g-2\Delta^*}{2N^{f_x}}\right)\partial_{p_f}\partial_y\rho_c.
\end{equation}
Note that the first two terms are $O(N^{0})$ and will thus equilibrate first. This determines the steady state probability distribution of $p_f$, $\sim e^{-2\gamma p_f^2/g}$. Interpreting the right-hand side of the master equation as an operator in phase space, we apply a Schrieffer-Wolff transformation to get rid of the term $\propto N^{-f_x}$, and then project the operator onto the steady state manifold of the $N^0$ term (operationally this means that we write $\rho_c\propto e^{-2\gamma p_f^2/g}P(y)$, apply the operator and integrate over $p$). This leads to an effective classical master equation for $y$
\begin{equation}
    \partial_tP(y)=\frac{g}{4N^{2f_x}}\partial_y^2P(y).
\end{equation}
This diffusion equation will be stabilized by the nonlinearity, which we now analyze. The nonlinearity comes from the $\hat{J}_x^2$ term in the Hamiltonian. Expressing the bosonic operators in terms of $\hat{x}_s$ and $\hat{p}_f$ we get that
\begin{equation}
    \frac{g\hat{J}_x^2}{N}=\frac{g\hat{x}^2}{2}-\frac{g}{16N}(u^4+1)\hat{x}_s^4+\frac{g u^4}{8N}(\hat{x}_s^3\hat{p}_f+\hat{p}_f\hat{x}_s^3),
\end{equation}
and we have kept up to terms with $\hat{x}_s^3$. The nonlinear terms induce the following evolution
\begin{align}\begin{split}
    -i\left[\frac{g}{16N}(u^4+1)\hat{x}_s^4,\hat{\rho}\right]&\to\frac{g N^{3f_x}}{4N}(u^4+1)y^3\partial_{p_f}\rho_c\\[5pt]
    -i\left[-\frac{g u^4}{8N}(\hat{x}_s^3\hat{p}_f+\hat{p}_f\hat{x}_s^3),\hat{\rho}\right]&\to-\frac{3gu^4 y^2 p_fN^{2f_x}}{4N}\partial_{p_f}\rho_c+\frac{gu^4N^{2f_x}}{4N}y^3\partial_y\rho_c    
\end{split}\end{align}
Projecting onto the steady state manifold of the $N^0$ term (and doing a Schrieffer-Wolff transformation to get rid of the first term) leads to the following master equation 
\begin{equation}
    \partial_tP=\frac{3gu^4N^{2f_x}}{4N}y^2P+\frac{gu^4 N^{2f_x}}{4N}y^3\partial_yP+\frac{g}{4N^{2f_x}}\partial_y^2P.
\end{equation}
We choose $f_x=1/4$ to arrive at
\begin{equation}
    \partial_tP=\frac{g}{N^{1/2}}\left[\frac{(\Delta^*)^2}{\gamma^2}\partial_y(y^3P)+\frac{1}{4}\partial_y^2P\right].
\end{equation}
Time evolution of the slow mode is thus reduced by a factor of $N^{-1/2}$. The relation between the $\hat{p}_f,\hat{x}_s$, and the original spin operators is
\begin{align}
    \begin{split}
        \hat{J}_x&=\frac{u}{2}(N^{3/4}\hat{y}-N^{1/2}\hat{p}_f)=\sqrt{\frac{\Delta^*}{2\gamma}}\big(N^{3/4}\hat{y}-N^{1/2}\hat{p}_f\big)\\[5pt]
        \hat{J}_y&=\frac{u^{-1}}{2}(N^{3/4}\hat{y}+N^{1/2}\hat{p}_f)=\frac{1}{2}\sqrt{\frac{\gamma}{2\Delta^*}}\big(N^{3/4}\hat{y}+N^{1/2}\hat{p}_f\big)
    \end{split}
\end{align}

\section{Effective Hamiltonian for the thermal phase transition of the Dicke model}\label{app:Thermal}

Here we derive the effective Hamiltonian that describes the thermal properties of the Dicke model in the vicinity of its phase transitions, Eq.~(\ref{eqn:Thermal:EffectiveHCriticalPoint}). We begin from
\begin{equation}
    \hat{H}_D=\omega\hat{c}^\dagger\hat{c}+\omega_0\hat{J}_z+\frac{2\lambda}{\sqrt{N}}\hat{J}_x(\hat{c}+\hat{c}^\dagger).
\end{equation}
The critical point is determined by $\lambda\sqrt{j}=\sqrt{\omega\omega_0}/2$ and $j=\tanh(\beta_c\omega_0/2)$. The quadratic approximation in the disordered phase at the critical point (including the degeneracy factor) is
\begin{equation}
    \frac{\hat{K}_{\text{eff}}}{\beta_c}=\omega\hat{c}^\dagger\hat{c}+\omega_0\Adag\Aop+\frac{\sqrt{\omega\omega_0}}{2}(\Aop+\Adag)(\hat{c}+\hat{c}^\dagger)+\frac{f''(j)\hat{l}^2}{\beta_c},
\end{equation} 
Omitting temporarily the $\hat{l}^2$ contribution, this model is more easily solved if we represent it in terms of quadratures $\hat{x}=(\Aop+\Adag)/\sqrt{2}$, $\hat{p}=-i(\Aop-\Adag)/\sqrt{2}$, $\hat{y}=(\hat{c}+\hat{c}^\dagger)/\sqrt{2}$ and $\hat{q}=-i(\hat{c}-\hat{c}^\dagger)/\sqrt{2}$, leading to
\begin{equation}
    \frac{\hat{K}_{\text{eff}}}{\beta_c}=-\frac{(\omega+\omega_0)}{2}+\frac{1}{2}\left(\omega\hat{q}^2+\omega_0\hat{p}^2\right)+\frac{1}{2}(\omega\hat{y}^2+\omega_0\hat{x}^2+2\sqrt{\omega\omega_0}\hat{x}\hat{y}).
\end{equation}
We first introduce the canonical rescalings $\tilde{p}=(\omega_0/\omega)^{1/4}\hat{p}$, $\tilde{x}=(\omega/\omega_0)^{1/4}\hat{x}$, and $\tilde{q}=(\omega/\omega_0)^{1/4}\hat{q}$, $\tilde{y}=(\omega_0/\omega)^{1/4}\hat{y}$ which makes uniform the terms quadratic in $\hat{p},\hat{q}$

\begin{equation}
    \frac{\hat{K}_{\text{eff}}}{\beta_c}=-\frac{(\omega+\omega_0)}{2}+\frac{\sqrt{\omega\omega_0}}{2}\left(\tilde{q}^2+\tilde{p}^2\right)+\frac{1}{2\sqrt{\omega\omega_0}}\left(\omega_0\tilde{x}+\omega\tilde{y}\right)^2.
\end{equation}
This representation makes it clear that the mode $\hat{k}=(\omega_0\tilde{x}+\omega\tilde{y})/(\omega_0^2+\omega^2)^{1/2}$ is gapped, while the mode $\hat{m}=(\omega\tilde{x}-\omega_0\tilde{y})/(\omega_0^2+\omega^2)^{1/2}$ is gapless
\begin{equation}
    \frac{\hat{K}_{\text{eff}}}{\beta_c}=-\frac{(\omega+\omega_0)}{2}+\frac{\sqrt{\omega\omega_0}}{2}\left(\tilde{p}_k^2+\tilde{p}_m^2\right)+\frac{(\omega_0^2+\omega^2)}{2\sqrt{\omega\omega_0}}\hat{k}^2,
\end{equation}
where $\hat{p}_{k,m}$ are the associated canonical momenta. The gapped mode can be put in to standard form by canonically rescaling $\hat{g}=(\omega_0^2+\omega^2)^{1/4}(\omega_0\omega)^{-1/4}\hat{k}$  and $\hat{p}_g= (\omega_0^2+\omega^2)^{1/4}(\omega_0\omega)^{-1/4}\hat{p}_k$, with $[\hat{g},\hat{p}_g]=i$. Furthermore, the quantity that fluctuates strongly at the critical point is $\hat{m}$ so schematically we have that $\hat{m}\gg\hat{k},\hat{p}_k,\hat{p}_m$ and therefore $\tilde{x}\approx\omega\,\hat{m}/(\omega_0^2+\omega^2)^{1/2}$, $\tilde{y}\approx -\omega_0\,\hat{m}/(\omega_0^2+\omega^2)^{1/2}$. We now add the nonlinearity, coming from the next term in the expansion of $\hat{J}_x$
\begin{equation}
    \hat{J}_x-\sqrt{\frac{Nj}{2}}\hat{x}\approx-\frac{1}{4\sqrt{2Nj}}\hat{x}^3+\frac{\hat{l}\hat{x}}{\sqrt{2j}}\approx-\frac{1}{4\sqrt{2Nj}}\left(\frac{\omega_0}{\omega}\right)^{3/4}\frac{\omega^3\hat{m}}{(\omega_0^2+\omega^2)^{3/2}}+\frac{\hat{l}\hat{m}}{\sqrt{2j}}\left(\frac{\omega_0}{\omega}\right)^{1/4}\frac{\omega}{(\omega_0^2+\omega^2)^{1/2}}
\end{equation}
The correction to the Hamiltonian from the nonlinearity is then
\begin{equation}
    \frac{\delta\hat{K}_{\text{eff}}^1}{\beta_c}=\frac{\sqrt{\omega\omega_0}}{\sqrt{Nj}}\left(\hat{J}_x-\sqrt{\frac{Nj}{2}}\hat{x}\right)(\hat{c}+\hat{c}^\dagger)=\frac{1}{4Nj}\frac{\omega^3\omega_0^2\,\hat{m}^4}{(\omega_0^2+\omega^2)^2}-\frac{1}{j\sqrt{N}}\frac{(\omega\omega_0)^{3/2}\hat{l}\hat{m}^2}{(\omega_0^2+\omega^2)}
\end{equation}
Note that the mode $\hat{m}$ couples to spin length fluctuations. To account for small deviations about the transition temperature, we recall that the effective Hamiltonian is obtained by adding the degeneracy factor to $\hat{H}_D$
\begin{equation}
    \hat{K}=\beta\hat{H}_D-\log(d_{\hat{J}}),
\end{equation}
so that a change in temperature is accounted for by
\begin{equation}
    \delta\hat{K}_{\text{eff}}^2=(\beta-\beta_c)\hat{H}_D.
\end{equation}
The most important terms that will be added are
\begin{equation}
   \delta\hat{K}_{\text{eff}}^2= -(\beta-\beta_c)\frac{\omega_0Nj}{2}-\omega_0(\beta-\beta_c)\sqrt{N}\hat{l}.
\end{equation}
The first term is a c-number but it may contribute to quantities like the average energy. Putting all these things together leads to
\begin{align}\begin{split}
    \hat{K}_{\text{eff}}^{\text{tr}}&=-\frac{\beta_c(\omega+\omega_0)}{2}+\frac{\beta_c(\omega_0^2+\omega^2)^{1/2}}{2}(\hat{p}_g^2+\hat{g}^2)\\
    &+\frac{\beta_c\sqrt{\omega\omega_0}}{2}\hat{p}_m^2+f''(j)\hat{l}^2-\omega_0(\beta-\beta_c)\sqrt{N}\hat{l}+\frac{\beta_c}{4Nj}\left[\frac{\omega^3\omega_0^2}{(\omega_0^2+\omega^2)^2}\right]\hat{m}^4-\frac{\beta_c}{j\sqrt{N}}\left[\frac{(\omega\omega_0)^{3/2}}{(\omega_0^2+\omega^2)}\right]\hat{l} \hat{m}^2
\end{split}\end{align}
If we were looking at ground state physics, we would scale $\hat{m}$ and $\hat{p}_m$ such that they would have the same $N$ prefactor and then adapt the scaling of $\lambda\sqrt{j}-\sqrt{\omega\omega_0}/2$ accordingly. This would lead to a quantum Hamiltonian with a gap $\propto N^{-1/3}$. At $T_c$ and large $N$, this nonlinear Hamiltonian would be highly excited, with an excitation level that depends on $N$ and would thus change the scalings that are relevant near the ground state phase transition. For the gapless mode, the correct procedure is to stabilize the thermal excitation of the system against the nonlinearity $\propto \hat{m}^4/N$ (by demanding that $\hat{K}_{\text{eff}}^{\text{tr}}\sim 1$) and to treat the mode classically. The effective Hamiltonian will thus have a quantum piece, coming from the gapped mode, and a classical piece, coming from the gapless mode and spin length fluctuations. To take the classical limit appropriately we define $\hat{s}$ and $\hat{p}_s$ such that 
\begin{align}
    \begin{split}
        \hat{s}&=\hat{m}\left[\frac{\beta_c^{1/4}\omega^{3/4}\omega_0^{1/2}}{ (\omega_0^2+\omega^2)^{1/2}(Nj)^{1/4}}\right]\\[10pt]
        \hat{p}_s&=\hat{p}_m\left[\frac{ (\omega_0^2+\omega^2)^{1/2}(j)^{1/4}}{\beta_c^{1/4}\omega^{3/4}\omega_0^{1/2}}\right].
    \end{split}
\end{align}
With these definitions, $[\hat{s},\hat{p}_s]=i N^{-1/4}$ and we can take the classical limit by letting commutators become Poisson brackets according to $\{\,\,,\,\}^{\text{pb}}\approx -iN^{-1/4}[\,\,,\,]$. The effective Hamiltonian is then
\begin{align}\begin{split}
    \hat{K}_{\text{eff}}^{\text{tr}}&=-\frac{\beta_c(\omega+\omega_0)}{2}+\frac{\beta_c(\omega_0^2+\omega^2)^{1/2}}{2}(\hat{p}_g^2+\hat{g}^2)\\[5pt]
    &+\frac{(\beta_c\omega_0)^{3/2}}{2\sqrt{j}}\left(\frac{\omega^2}{\omega^2+\omega_0^2}\right)p_s^2+f''(j)l^2+\frac{s^4}{4}+\left[\sqrt{N}(\beta-\beta_c)\omega_0-\sqrt{\frac{\beta_c\omega_0}{j}} \,s^2\right]l
\end{split}\end{align}
where we are now treating $s,p_s,l$ as classical variables. The last thing to do is to scale the distance to the critical point with $N$ such that
\begin{equation}
    \xi=\frac{(\beta-\beta_c)\sqrt{N}}{\beta_c}
\end{equation}
so that the effective Hamiltonian reads
\begin{align}\begin{split}
   \hat{K}_{\text{eff}}^{\text{tr}}&=-\frac{\beta_c(\omega+\omega_0)}{2}+\frac{\beta_c(\omega_0^2+\omega^2)^{1/2}}{2}(\hat{p}_g^2+\hat{g}^2)\\[5pt]
    &+\frac{(\beta_c\omega_0)^{3/2}}{2\sqrt{j}}\left(\frac{\omega^2}{\omega^2+\omega_0^2}\right)p_s^2+f''(j)l^2+\frac{s^4}{4}+\left[\xi\beta_c\omega_0-\sqrt{\frac{\beta_c\omega_0}{j}} \,s^2\right]l
\end{split}\end{align}
For completeness, we include here the exact relation between the original transverse bosons $\Aop,\Adag$ and the final expression in terms of $\hat{g},\hat{s}$:
\begin{align}\label{eqn:app:Thermal:BosonRelations}
    \begin{split}
        \hat{g}&=\left(\frac{\omega_0^2+\omega^2}{\omega_0\omega}\right)^{1/4}\left[\frac{\omega_0(\omega/\omega_0)^{1/4}(\Aop+\Adag)+\omega(\omega_0/\omega)^{1/4}(\hat{c}+\hat{c}^\dagger)}{\sqrt{2}(\omega_0^2+\omega^2)^{1/2}}\right]\\[5pt]
        \hat{p}_g&=\left(\frac{\omega_0^2+\omega^2}{\omega_0\omega}\right)^{-1/4}\left[\frac{\omega_0(\omega_0/\omega)^{1/4}(\Aop-\Adag)+\omega(\omega/\omega_0)^{1/4}(\hat{c}-\hat{c}^\dagger)}{i\sqrt{2}(\omega_0^2+\omega^2)^{1/2}}\right]\\[10pt]
        \hat{s}&=\left[\frac{\beta_c^{1/4}\omega^{3/4}\omega_0^{1/2}}{ (\omega_0^2+\omega^2)^{1/2}(Nj)^{1/4}}\right]\left[\frac{\omega(\omega/\omega_0)^{1/4}(\Aop+\Adag)-\omega_0(\omega_0/\omega)^{1/4}(\hat{c}+\hat{c}^\dagger)}{\sqrt{2}(\omega_0^2+\omega^2)^{1/2}}\right]\\[5pt]
        \hat{p}_s&=\left[\frac{ (\omega_0^2+\omega^2)^{1/2}(j)^{1/4}}{\beta_c^{1/4}\omega^{3/4}\omega_0^{1/2}}\right]\left[\frac{\omega(\omega_0/\omega)^{1/4}(\Aop-\Adag)-\omega_0(\omega/\omega_0)^{1/4}(\hat{c}-\hat{c}^\dagger)}{i\sqrt{2}(\omega_0^2+\omega^2)^{1/2}}\right]
    \end{split}
\end{align}

\end{document}